\documentclass[journal,twoside,web]{ieeecolor}
\usepackage{generic}
\usepackage{cite}
\usepackage{amsmath,amssymb,amsfonts}
\usepackage{algorithmic}
\usepackage{graphicx}
\usepackage{textcomp}
\usepackage[caption=false,font=scriptsize,labelfont=sf]{subfig}
\usepackage{babel}
\usepackage{array}
\usepackage{textcomp}
\usepackage{stfloats}
\usepackage{xcolor}
\usepackage{booktabs}
\usepackage[export]{adjustbox}
\usepackage{url}
\usepackage{verbatim}
\usepackage{bm}
\usepackage{accents}
\usepackage{multicol}
\usepackage{multirow}
\usepackage{eqparbox}
\usepackage{enumerate}
\usepackage{nccmath}
\usepackage{nomencl}
\makenomenclature
\usepackage{soul}
\usepackage{makecell}
\usepackage{bbding}
\usepackage{etoolbox}
\renewcommand\nomgroup[1]{%
  \item[\bfseries
  \ifstrequal{#1}{A}{Indices}{%
  \ifstrequal{#1}{B}{Parameters}{%
  \ifstrequal{#1}{C}{Variables}{}}}%
]}

\newcolumntype{M}[1]{>{\centering\arraybackslash}m{#1}}
\newcolumntype{P}[1]{>{\centering\arraybackslash}p{#1}}

\def\BibTeX{{\rm B\kern-.05em{\sc i\kern-.025em b}\kern-.08em
    T\kern-.1667em\lower.7ex\hbox{E}\kern-.125emX}}
\begin{document}

\title{Attentive Convolutional Deep Reinforcement Learning for Optimizing Solar-Storage Systems in Real-Time Electricity Markets}

\author{Jinhao Li, Changlong Wang, Hao Wang
\thanks{This work was supported in part by the Australian Research Council (ARC) Discovery Early Career Researcher Award (DECRA) under Grant DE230100046. (Corresponding author: Hao Wang.)}
\thanks{J. Li and H. Wang are with the Department of Data Science and AI, Faculty of IT and Monash Energy Institute, Monash University, Melbourne, VIC 3800, Australia (e-mail: \{jinhao.li,hao.wang2\}@monash.edu).}
\thanks{C. Wang is with the Department of Civil Engineering, Monash University, Melbourne, VIC 3800, Australia (e-mail: chang.wang@monash.edu).}}

\maketitle

\begin{abstract}
This paper studies the synergy of solar-battery energy storage system (BESS) and develops a viable strategy for the BESS to unlock its economic potential by serving as a backup to reduce solar curtailments while also participating in the electricity market. 
We model the real-time bidding of the solar-battery system as two Markov decision processes for the solar farm and the BESS, respectively. We develop a novel deep reinforcement learning (DRL) algorithm to solve the problem by leveraging attention mechanism (AC) and multi-grained feature convolution to process DRL input for better bidding decisions. Simulation results demonstrate that our AC-DRL outperforms two optimization-based and one DRL-based benchmarks by generating $23\%$, $20\%$, and $11\%$ higher revenue, as well as improving curtailment responses. The excess solar generation can effectively charge the BESS to bid in the market, significantly reducing solar curtailments by $76\%$ and creating synergy for the solar-battery system to be more viable.
\end{abstract}


\nomenclature[A]{$t$}{Index of time frame}

\nomenclature[B,01]{$\Delta t$}{NEM dispatch interval (mins)}
\nomenclature[B,02]{$T$}{Operation horizon}
\nomenclature[B,03]{$H$}{Period length for battery degradation calculation}
\nomenclature[B,04]{$c$}{Battery cost (AU\$/MWh)}
\nomenclature[B,05]{$\alpha$}{Penalty coefficient of solar bid}
\nomenclature[B,06]{$k_t^\text{deg}$}{Coefficient associated with battery calendar aging and cycle aging}
\nomenclature[B,07]{$\sigma$}{Transmission limit coefficient}
\nomenclature[B,08]{$\eta^\text{Dch}$}{BESS discharging efficiency}
\nomenclature[B,09]{$\eta^\text{Ch}$}{BESS charging efficiency}
\nomenclature[B,10]{$P_\text{max}^\text{Bat}$}{BESS rated power (MW)}
\nomenclature[B,11]{$P_\text{max}^\text{S}$}{Solar farm installed capacity (MW)}
\nomenclature[B,12]{$E_\text{min}^\text{Bat}$}{BESS lower energy limit (MWh)}
\nomenclature[B,13]{$E_{t,\text{max}}^\text{Bat}$}{BESS upper energy limit at time frame $t$ (MWh)}
\nomenclature[B,14]{$\lambda_t$}{Spot price at time frame $t$ (AU\$/MWh)}
\nomenclature[B,15]{$e_t$}{BESS capacity at time frame $t$ (MWh)}
\nomenclature[B,16]{$d_t^\text{Deg}$}{BESS degradation coefficient}
\nomenclature[B,17]{$p_t^\text{S,Ava}$}{Solar plant availability (MW)}

\nomenclature[C,01]{$p_t^\text{S}$}{Solar bid power in the spot market (MW)}
\nomenclature[C,02]{$p_t^\text{S,Act}$}{Actual solar PV generation in the spot market (MW)}
\nomenclature[C,03]{$p_t^\text{S,Dis}$}{Dispatched solar PV generation in the spot market (MW)}
\nomenclature[C,04]{$v_t^\text{Dch}$}{Binary variable indicating the BESS discharge operation}
\nomenclature[C,05]{$v_t^\text{Ch}$}{Binary variable indicating the BESS charge operation}
\nomenclature[C,06]{$p_t^\text{Bat,SM}$}{BESS bid power in the spot market (MW)}
\nomenclature[C,07]{$\hat{p}_t^\text{Bat,SC}$}{Power that BESS plans to draw from solar curtailments (MW)}
\nomenclature[C,08]{$p_t^\text{S,SC}$}{Curtailed solar PV generation if there is no BESS for curtailment reduction (MW)}
\nomenclature[C,09]{$p_t^\text{Bat,SC}$}{Actual power drawn by BESS from the otherwise curtailed solar PV power (MW)}
\nomenclature[C,10]{$\Delta e_t$}{BESS capacity change at time frame $t$ (MWh)}

\printnomenclature

\section{Introduction} \label{sec:intro}
\subsection{Background and Literature Review}Solar photovoltaic (PV) has attracted nearly half of the global renewable investment and become the leading source of utility-scale renewables~\cite{comello2019}. 
However, due to the inherent variability of solar PV generation, curtailment of solar PV outputs is often inevitable but causes significant economic losses for solar farms~\cite{mallapragada2020}.
Therefore, effective efforts for solar curtailment mitigation are urgent for power system operators as a key factor for the successful grid transition, as well as for solar farm owners for profitability concerns.

The grid-scale battery energy storage system (BESS) has emerged as a critical solution for reducing solar curtailments and improving the economic performance of solar farms~\cite{bumpus2017}. In co-located solar-battery systems, the BESS usually serves as the onsite backup source to smooth the fluctuating solar generation by absorbing excess solar energy (which is otherwise curtailed) and releasing stored energy when the solar PV generation is insufficient. However, the limited ancillary role fails to unlock the BESS's economic potential and cannot justify the economic viability of co-located BESS, due to the high upfront cost of the BESS despite its declining cost over the past decades. To enhance the economic viability, the BESS can take advantage of its operational flexibility to perform energy arbitrage (i.e., buy low and sell high) in the wholesale electricity market as a price-taker, creating an additional revenue stream for the BESS complementing its ancillary role in reducing solar curtailment. Therefore, it is of great significance to design an effective coordination strategy for the co-located solar-battery system to concurrently reduce solar curtailment and enhance the BESS's viability, creating a win-win situation for solar farms, the BESS, and the power system.
However, designing such an effective coordination strategy is not trivial but challenging due to the uncertainty of solar energy, volatility of the energy market, and BESS decision coupling between curtailment management and energy arbitrage.

Optimization-based methods have been introduced to optimize bidding decisions of the co-located solar-battery system in the electricity market. To tackle the uncertainties of solar generation and the electricity market, stochastic optimization (SO)~\cite{graca2023} and robust optimization (RO)~\cite{attarha2019} have been introduced to maximize the overall revenue of the solar-battery system, but often overlook the importance of solar curtailment, thereby failing to take advantage of curtailed solar energy to charge the BESS without additional fees. In addition, the above works primarily focused on the day-ahead electricity market, i.e., scheduling the BESS's operations for the next $24$ hours. In contrast to the day-ahead market, the real-time electricity market presents higher profitability than the day-ahead market~\cite{aemo2023}, but also greater challenges due to the high volatility of the real-time market and complex price drivers.

Optimization approaches for real-time bidding, in particular the model predictive control (MPC) and its variants, have been widely studied in the literature. For example, a deterministic MPC (DMPC) method has been proposed by~\cite{badoual2021_DMPC} for a BESS taking part in real-time electricity market. Stochastic MPC (SMPC)~\cite{conte2018_SMPC}, robust MPC (RMPC)~\cite{xie2021_RMPC}, and ensemble nonlinear MPC (EnNMPC)~\cite{li2023_EnNMPC} methods have also been introduced to further characterize and capture the uncertainties of both solar generation and electricity price. However, the aforementioned four kinds of MPC-based methods also depend on the quality of forecast results, as the predictions of solar generation and prices are the parameters for the optimization solver, thereby affecting the optimality of bidding decisions. Additionally, the electricity price is notoriously challenging to predict~\cite{weron2014}.

Besides MPC-based approaches, deep reinforcement learning (DRL) has drawn increasing attention in real-time decision-making problems, for its powerful capability to directly learn the dynamics of the environment (such as solar generation and the electricity market) and consequently capture the uncertainties of time-varying parameters without using explicit forecast models.  
Wang \textit{et al.}~\cite{wang2023_reviewer2} managed a multi-energy microgrid via a DRL-based approach, while they operated the BESS as an ancillary asset. One recent study~\cite{dolatabadi2023_reviewer6_oneEntity} discussed the real-time bidding of a co-located solar-battery system. However, the integrated system functioned as a single participant in real-time markets, ignoring the underlying coordination between the two facilities.

The synergy of a co-located wind-battery system in real-time markets has been studied in~\cite{li2023}, but one essential factor was neglected. How to effectively use available information as features for deep reinforcement learning (DRL) is not well understood, but it is essential to DRL-based bidding decision-making. For example, the BESS's arbitrage decision is influenced by the energy price, and energy exchange in the market is impacted by the physical battery capacity. Existing studies have not addressed this question, thus highlighting the need for research on effectively harnessing input features, including solar generation, battery capacity, and energy price, to support DRL-based bidding strategies.

\subsection{Contributions and Paper Structure}
Motivated by the need for effective coordination strategies for the solar-battery system, encompassing curtailment mitigation and market participation (i.e., real-time energy arbitrage), we leverage the data-driven ability of DRL to develop a novel DRL-based bidding strategy for co-located solar-battery systems. Our strategy aims to concurrently manage solar curtailment and optimize the system's participation in the wholesale real-time market. By employing DRL, we can learn the uncertainties of solar power and energy prices in a model-free manner and derive the coordination strategy from historical experiences through interactive training without relying on prior knowledge or forecasts. Furthermore, to better explore how various input features influence the bidding decision-making for both the solar farm and the BESS, we propose a novel DRL network structure comprising attention mechanism and multi-grained feature convolution. Specifically, the attention mechanism can exploit the correlations among features and emphasize the relative significance of each feature, which is then processed by multi-grained feature convolution to make bidding decisions. This structure enables both the BESS and the solar farm to focus on different input features during the dynamic bidding process. We refer to our method as AC-DRL (\textbf{A}ttentive \textbf{C}onvolutional \textbf{DRL}). 
The main contributions of this paper are summarized as follows.
\begin{itemize}
    \item \textit{Optimizing Coordination of Solar Curtailment Reduction and BESS Bidding in Real-Time Electricity Market:} We study coordinated solar-battery operations to simultaneously manage solar curtailment and perform energy arbitrage in a real-time market, creating a more promising revenue stream than purely operating as onsite backup sources. Our work provides a viable case for the co-location of renewable generators and the BESS by demonstrating the synergy of the co-located solar-battery system to unlock its economic potential.
    \item \textit{Attentive Convolutional DRL-based Bidding:} We design a novel DRL network structure, namely AC-DRL, incorporating a stacked attention mechanism and a multi-grained feature convolution module. Our AC-DRL fully explores the correlations of DRL input and effectively identifies the relative significance of input features, resulting in better and more informed decision-making. With the AC-DRL, we further decouple the market participation of the co-located system into two correlated Markov decision processes, enabling us to better analyze how multiple factors, e.g., energy price and solar generation, affect the decision-making of the co-located system.
    \item \textit{Simulations and Insights Based on Real-World Data}: Using realistic solar farm data collected from the Australian National Electricity Market (NEM), our simulations demonstrate the effectiveness of our AC-DRL method. The results show that AC-DRL significantly outperforms both MPC-based and DRL-based benchmarks. It is revealed that effective solar curtailment management is the key to the successful coordination of the two assets, improving energy arbitrage and enhancing the economic viability of the co-located solar-battery system.
\end{itemize}

The remainder of this paper is organized as follows. Section \ref{sec:system_model} introduces the solar-battery system and formulates the coordination problem of its participation in real-time market. Section \ref{sec:method} disentangles the system's bidding process and presents our AC-DRL framework to optimize the overall revenue while reducing solar curtailments. Section \ref{sec:exps} presents simulation results, and Section \ref{sec:conclusion} concludes this paper.

\begin{figure}[!t]
    \centering\includegraphics[width=.75\linewidth]{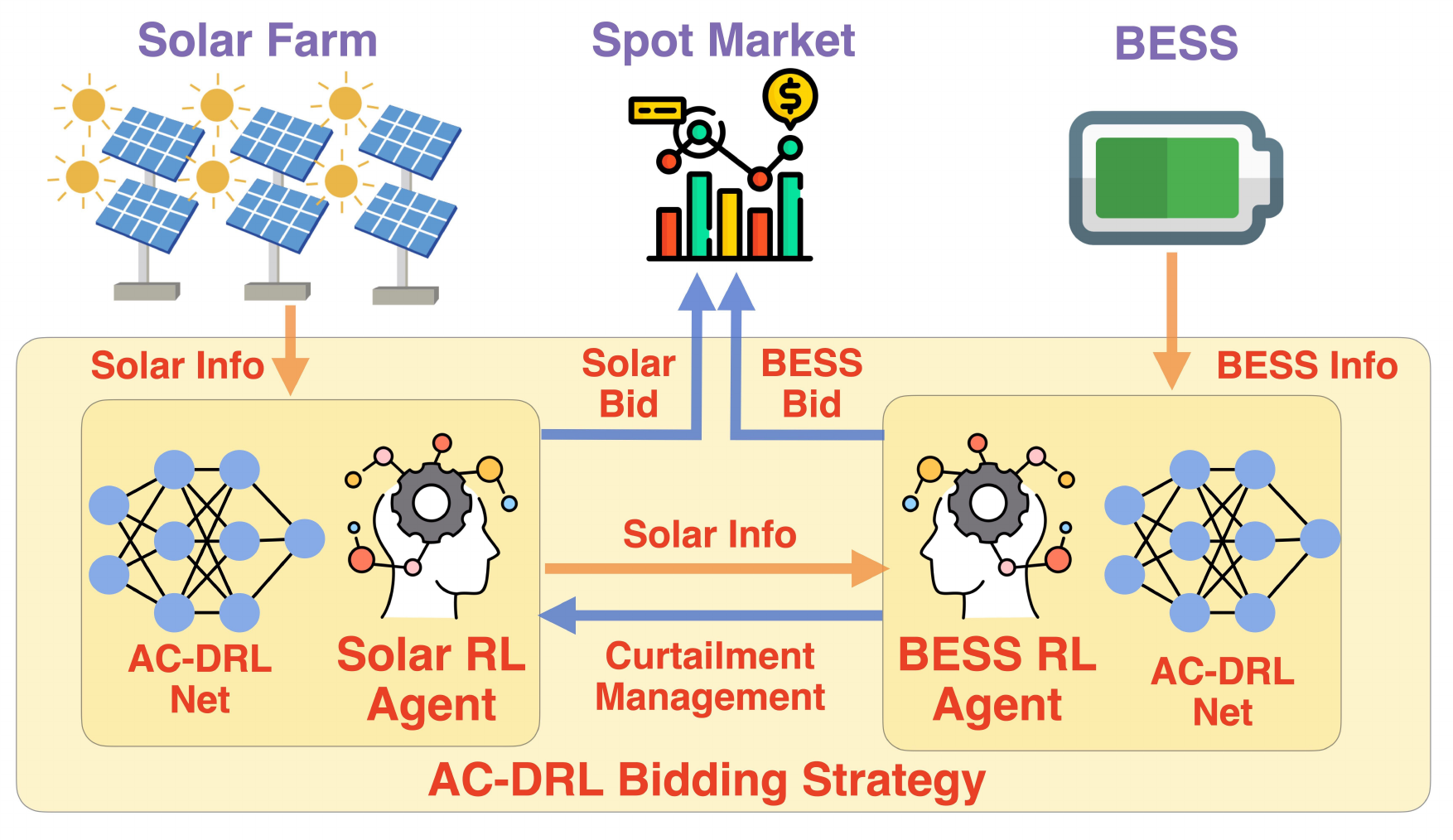}
    \caption{The system model paradigm.}
    \label{fig:system_model}
\end{figure}

\section{System Model} \label{sec:system_model}
We consider a co-located solar-battery system participating as a price-taker in real-time spot market (Fig. \ref{fig:system_model}). This assumption is reasonable for small systems with negligible market power, as distributed solar and storage resources proliferate. Our system size is minor compared to the significant NEM demand~\cite{aemo2020}.
The price-taker focus enables studying the key problem of solar-battery coordination and bidding agent design, while isolating broader market effects. It has precedent in DRL~\cite{zhu2023_reviewer2} and optimization~\cite{graca2023} bidding literature. Although simplified compared to full multi-agent interactions, it allows initial benchmarking of bidding performance. Nevertheless, there is still an export limit for the co-located system to the grid, which is assumed to be $62.5\%$ of the solar-battery system's installed capacity, as described in~\cite{daggett2017}. In the co-located solar-battery system, the BESS simultaneously performs energy arbitrage in the spot market and charges otherwise curtailed solar energy from the onsite solar farm during curtailment events. This section is organized as follows. The background on the spot market is introduced in Section \ref{subsec:system_model_preliminary}. In Section \ref{subsec:system_model_revenue}, we model the solar farm and the BESS revenue streams under various operational conditions. The overall bidding problem of the solar-battery system is formulated in Section \ref{subsec:system_model_formulation}.

\subsection{Background of the Spot Market} \label{subsec:system_model_preliminary}
The spot market, as the main component of the NEM, functions as a real-time market for wholesale electricity transactions, where the power mismatches between generators and loads are immediately balanced by the Australian Energy Market Operator (AEMO) through a centrally coordinated dispatch process~\cite{aemo2020}. Specifically, AEMO receives bids from generators every five minutes, and dispatches generators in a cost-effective manner by ranking their bids from low to high, forming a bidding stack. The bid price that meets the final demand determines the market clearing price, known as the spot price (in AU\$/MWh)~\cite{aemo2020}. Our solar-battery system, like generators, once dispatched, will be paid at the spot price.

\subsection{The Solar-Battery System and Revenue Streams} \label{subsec:system_model_revenue}
\subsubsection{Solar Farm} \label{subsubsec:system_model_revenue_solar}
Due to the uncertainty of solar generation, most solar farms in the NEM register as semi-scheduled generators. The AEMO requires the solar farm to constantly update its plant availability (also known as the upper power limit), denoted as $p_t^\text{S,Ava}$ from onsite monitoring devices~\cite{aemo2022}, based on which the solar farm can schedule a dispatch target, i.e., the amount of bid power (in MW) in the spot market, denoted as $p_t^\text{S}$, to fulfill in the next dispatch interval.

The intermittence of solar PV power often results in deviations between the dispatch target and the actual solar generation, denoted as $p_t^\text{S,Act}$. The deviation can make it challenging for the solar farm to deliver the exact amount of dispatch target. For instance, if there is a solar shortage, i.e., $p_t^\text{S,Act}<p_t^\text{S}$, the dispatch target can only be partially met, while the excessive solar output will be curtailed to match the dispatch target if there is excess solar generation, i.e., $p_t^\text{S,Act}>p_t^\text{S}$. Therefore, the actual dispatched solar power can be defined as $p_t^\text{S,Dis}=\min\{p_t^\text{S,Act},p_t^\text{S}\}$. Moreover, to regulate the bidding behaviors of the solar farm to comply with the market rules~\cite{aemo2022}, we introduce a penalty term if the solar farm does not meet the dispatch target. We define $\lambda_t$ as the spot price, $T$ as the overall time frame, $\Delta t$ as the NEM dispatch interval (in $5$-minute resolution), and $\alpha$ as the aforementioned penalty coefficient~\cite{xie2021_RMPC}. The revenue stream of the solar farm from the spot market can thus be formulated as
\begin{equation}
    \label{eq:revenue_solar}
    R^\text{S} = \Delta t\sum_{t=1}^T \lambda_t\left( p_t^\text{S,Dis}-\alpha|p_t^\text{S,Dis}-p_t^\text{S}|\right).
\end{equation}

\subsubsection{BESS} \label{subsubsec:system_model_revenue_BESS}
Price fluctuations in the spot market serve as a reflection of its inherent volatility, where the lack of generation or demand leads to an increase or decrease in prices, respectively. The BESS can take advantage of both the market stochasticity and its operational flexibility by switching between its two working modes (storage and generation) to buy energy at low prices and sell it at high prices, known as energy arbitrage in the spot market. Given that the BESS cannot simultaneously charge and discharge, we introduce two binary variables, $v_t^\text{Ch}$ and $v_t^\text{Dch}$, to indicate the BESS's operational mode, leading to the following constraints
\begin{equation}
    \label{eq:cons_BESS_ch_dch}
    v_t^\text{Dch} + v_t^\text{Ch} \leq 1, ~ v_t^\text{Dch},v_t^\text{Ch} \in \{0,1\}.
\end{equation}
The BESS sits idle when both binary variables are zero.

Let $p_t^\text{Bat,SM}$ be the BESS's bid power in the spot market. We define BESS's revenue from the spot market as
\begin{equation}
    \label{eq:revenue_BESS}
    R^\text{Bat} = \Delta t\sum_{t=1}^T\left(v_t^\text{Dch}-v_t^\text{Ch}\right)\lambda_t p_t^\text{Bat,SM}.
\end{equation}
In the co-located solar-battery system, besides power bought from the spot market, the BESS can also store the otherwise curtailed solar generation from the onsite solar farm. We denote the power planned to charge the BESS from solar curtailments as $\hat{p}_t^\text{Bat,SC}$. As discussed, the BESS cannot concurrently charge and discharge. Thus, the BESS cannot absorb curtailed solar power while exporting its own power to the spot market. Such an operational constraint can be formulated as
\begin{equation}
    v_t^\text{Dch}\hat{p}_t^\text{Bat,SC} = 0
\end{equation}

The onsite curtailed solar power can be expressed as
\begin{equation}
    \label{eq:solar_curtailed_power}
    p_t^\text{S,SC} = \left(p_t^\text{S,Act}-p_t^\text{S}\right)\mathbb{I}\left(p_t^\text{S,Act}>p_t^\text{S}\right),
\end{equation}
where $\mathbb{I}(p_t^\text{S,Act}>p_t^\text{S})$ indicates the occurrence of a solar curtailment event. Therefore, the actual power that the BESS draws from the onsite solar farm can be presented as
\begin{equation}
    p_t^\text{Bat,SC} = \min \{\hat{p}_t^\text{Bat,SC},p_t^\text{S,SC}\}.
\end{equation}

Moreover, frequent charge and discharge of the BESS leads to battery degradation. We introduce a non-linear battery degradation model proposed in~\cite{xu2018batterydeg}. Let $k^\text{deg}_t$ be the coefficient associated with battery calendar aging and cycle aging, which can be calculated using the rainflow cycle-counting algorithm~\cite{musallam2012rainflow}. After a period of charging and discharging operations (with a period length of $H$), the remaining battery storage capacity (i.e., the maximum energy that can be stored in the BESS) can be written as
\begin{equation}
    E_{{t+H},\text{max}}^\text{Bat}= E_{t,\text{max}}^\text{Bat}e^{-k^\text{deg}_t},
\end{equation}
where $E_{t,\text{max}}^\text{Bat}$ is the battery storage capacity before arbitrage operations. With the capacity decline, we further estimate the battery degradation coefficient, denoted by $d^\text{deg}_t$, for a given period of arbitrage operations~\mbox{\cite{cao2020batterydeg}}, which can be formulated as
\begin{equation}
    \label{eq:battery_deg_coeff}
    d^\text{Deg}_{t'} = \frac{c\left(E_{t',\text{max}}^\text{Bat}-E_{t'+H,\text{max}}^\text{Bat}\right)}{\Delta t\sum_{t=t'}^{t'+H} \left|p_t^\text{Bat,SM}+p_t^\text{Bat,SC}\right|},
\end{equation}
where $c$ is the battery cost per MWh~\mbox{\cite{cao2020batterydeg}}.
We therefore formulate the battery degradation cost as
\begin{equation}
    \label{eq:battery_deg_cost}
    C^\text{Bat} = \Delta t\sum_{t=1}^T d^\text{Deg}_t\left|p_t^\text{Bat,SM}+p_t^\text{Bat,SC}\right|.
\end{equation}
The degradation coefficient $d_t^\text{deg}$ is updated every $H$ intervals.

\subsection{Bidding Formulation of the Solar-Battery System} \label{subsec:system_model_formulation}
The market participation of the co-located solar-battery system can be formulated as an optimization problem, whose objective is to maximize the overall profit from the spot market, written as
\begin{equation}
    \label{eq:optimization_obj}
    \max \hspace{0.25em} R^\text{S} + R^\text{Bat} - C^\text{Bat},
\end{equation}
which includes the revenue of the solar farm, the revenue of the BESS through energy arbitrage in the spot market, and the battery degradation cost of the BESS. 

Real-time bidding of the solar farm and the BESS is constrained as
\begin{align}
    \label{eq:cons_solar_power}
    0&\leq p_t^\text{S}\leq p_t^\text{S,Ava},\\
    \label{eq:cons_BESS_Spot_power}
    0&\leq p_t^\text{Bat,SM}\leq P_\text{max}^\text{Bat},\\
    \label{eq:cons_BESS_SC_power}
    0&\leq \hat{p}_t^\text{Bat,SC}\leq P_\text{max}^\text{Bat},\\
    \label{eq:cons_BESS_tot_power}
    0&\leq p_t^\text{Bat,SM}+\hat{p}_t^\text{Bat,SC}\leq P_\text{max}^\text{Bat},\\
    \label{eq:cons_transmission}
    0&\leq p_t^\text{S} + p_t^\text{Bat,SM} + \hat{p}_t^\text{Bat,SC}\leq \sigma\left(P_\text{max}^
    \text{S}+P_\text{max}^\text{Bat}\right),
\end{align}
where $P_\text{max}^\text{Bat}$ and $P_\text{max}^\text{S}$ are the rated power of the BESS and the solar farm, respectively, and $\sigma$ is the transmission limit coefficient of the co-located solar-battery system, which is set as $62.5\%$ by default. Equation \eqref{eq:cons_solar_power} is the constraint for the solar bid power. Equations \eqref{eq:cons_BESS_Spot_power} and \eqref{eq:cons_BESS_SC_power} constrain the BESS bid power in the spot market and the power intended to draw from the onsite solar curtailment. Equation \eqref{eq:cons_BESS_tot_power} indicates that the sum of the market bid and absorbed power from the onsite solar farm cannot exceed the BESS's rated power. Equation \eqref{eq:cons_transmission} describes the export limit of the co-located system, subject to its installed capacity.

The BESS's capacity at each dispatch interval must be within its lower and upper energy limits denoted by $E_\text{min}^\text{Bat}$ and $E_{t,\text{max}}^\text{Bat}$, which can be formulated as
\begin{equation}
    \label{eq:cons_BESS_energy_change}
    E_\text{min}^\text{Bat} \leq e_{t-1} + \Delta e_t\leq E_{t,\text{max}}^\text{Bat},
\end{equation}
where $e_{t-1}$ is BESS's capacity after the last dispatch interval and $\Delta e_t$ is the current energy change. Note that the upper energy limit $E_{t,\text{max}}^\text{Bat}$ continuously decreases due to the battery degradation. The energy change is due to the power exchange in the spot market and curtailed solar energy absorption from the onsite solar farm, which can be further expressed as
\begin{equation}
     \hspace{-2mm}\Delta e_t = \Delta t \left[\left(v_t^\text{Ch}\eta^\text{Ch}-\frac{v_t^\text{Dch}}{\eta^\text{Dch}}\right)p_t^\text{Bat,SM} + v_t^\text{Ch}\eta^\text{Ch}p_t^\text{Bat,SC}\right],
\end{equation}
where $\eta^\text{Ch}$, $\eta^\text{Dch}$ are charging and discharging efficiencies of the BESS.

\begin{figure}[!t]
    \centering
    \includegraphics[width=.8\linewidth]{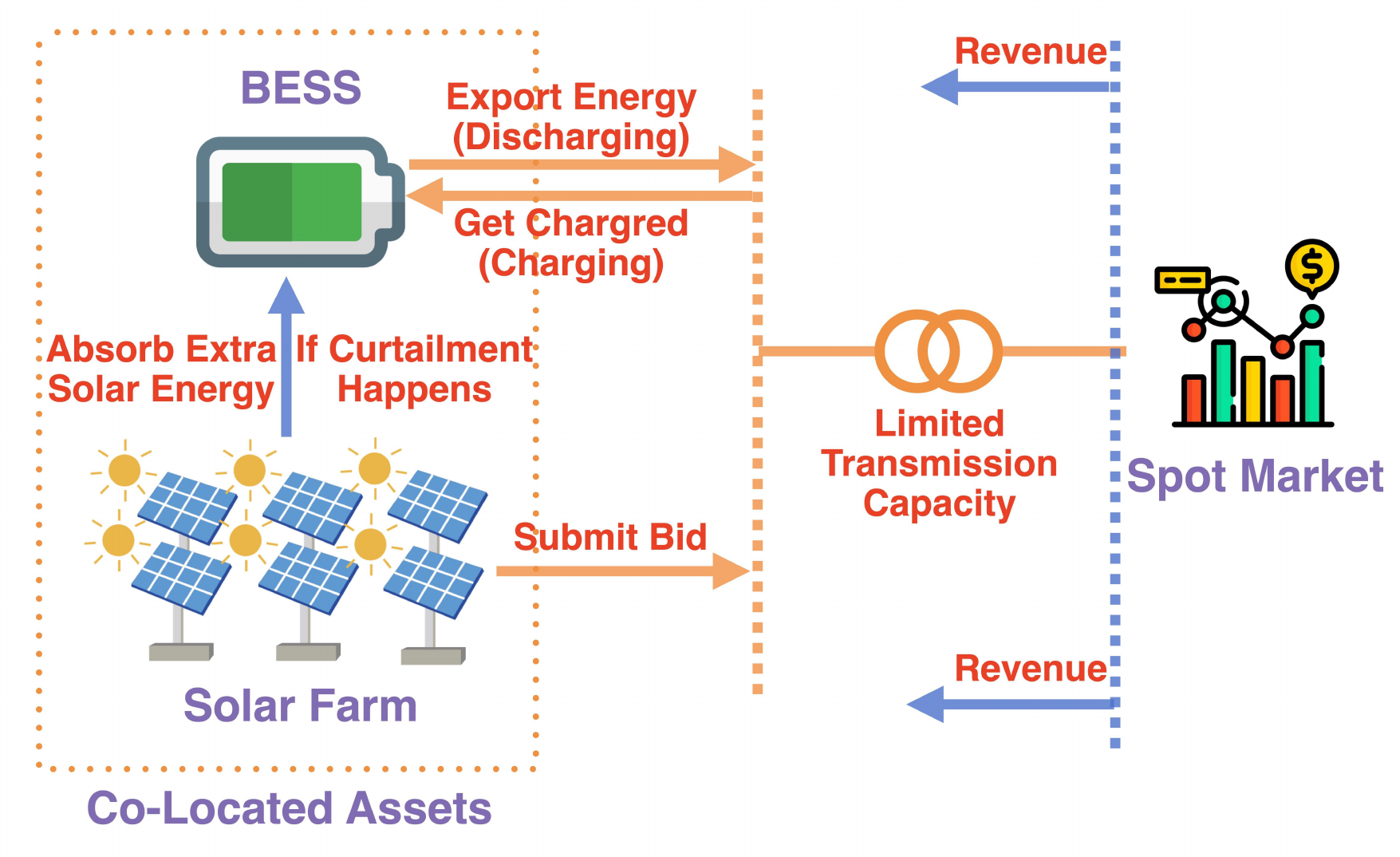}
    \caption{The detailed system configuration.}    \label{fig:detailed_system_config}
\end{figure}

The detailed system configuration of our co-located solar-battery system is illustrated in Fig.~\ref{fig:detailed_system_config}.

\section{Methodology} \label{sec:method}
To solve the optimization problem formulated in Section \ref{subsec:system_model_formulation} and better characterize the synergy of the co-located system, we disentangle the system's bidding process and model it into two Markov decision processes (MDP) for the solar farm and the BESS, respectively. We then present our AC-DRL network structure and introduce the deep deterministic policy gradient (DDPG) algorithm to jointly maximize the expected cumulative revenue of the solar-battery system in the spot market. This section is organized as follows. Section \ref{subsec:method_MDP} presents MDP formulation, and Section \ref{subsec:method_ACDRL} proposes our AC-DRL coordinated bidding strategy.

\subsection{MDP Modeling} \label{subsec:method_MDP}
The bidding strategy of the solar-battery system can be affected by multiple factors, e.g., solar PV generation, spot price, and BESS's capacity. To have a better understanding of the interaction between the solar farm and the BESS, we decouple the system's bidding process into two MDPs for the solar farm and the BESS, each of which consists of four elements: state space $\mathbb{S}^\text{S}/\mathbb{S}^\text{Bat}$, action space $\mathbb{A}^\text{S}/\mathbb{A}^\text{Bat}$, probability space $\mathbb{P}^\text{S}/\mathbb{P}^\text{Bat}$, and reward space $\mathbb{R}^\text{S}/\mathbb{R}^\text{Bat}$. Moreover, it is also reasonable to disentangle the system's bidding, as the solar farm and the BESS may belong to different entities. Additionally, a single DRL agent has been revealed to be inefficient in tackling multiple tasks (e.g., solar curtailment management and bidding of the two assets) with only one reward signal~\cite{bai2023_MTDRL}.

\textbf{State Space $\mathbb{S}$:} As discussed, the external factors (e.g., the spot price) and internal factors (e.g., the BESS's capacity and historical solar generation) should be considered as the state in the MDP models. Specifically, the state of the solar farm includes the last spot price $\lambda_{t-1}$, the last actual solar generation $p_{t-1}^\text{S,Act}$, the last solar power deviation defined as $p_{t-1}^\text{S,Dev}=p_{t-1}^\text{S,Act}-p_{t-1}^\text{S}$, and the current hour index $h_t\in\{0,\frac{1}{23},\frac{2}{23},\cdots,1\}$. Since there is no solar generation during night times, such an hour index can partially characterize the effect of solar irradiation on solar output profiles. For the BESS, its state consists of the last spot price, the last solar deviation, the hour index, and the BESS's capacity $e_{t-1}$. Moreover, to achieve effective solar curtailment management, we also consider the number of solar curtailment events (assuming there is no BESS for curtailment reduction) within the latest $L$ dispatch intervals, denoted by $f_{t-1}^\text{SC}$, in the BESS's state. Moreover, we include the average amount of solar curtailment (in MWh) in the last $L$ dispatch intervals, which is denoted as $m^\text{SC}_{t-1}$. The states of the solar farm and the BESS can be summarized as
\begin{align}
    \label{eq:MDP_state_solar}
    \bm{s}_t^\text{S} &= \left[ \lambda_{t-1},p_{t-1}^\text{S,Act},p_{t-1}^\text{S,Dev},h_t \right],\\
    \label{eq:MDP_state_BESS}
    \bm{s}_t^\text{Bat} &= \left[\lambda_{t-1},e_{t-1},p_{t-1}^\text{S,Dev},f_{t-1}^\text{SC},m^\text{SC}_{t-1},h_t\right].
\end{align}

\textbf{Action Space $\mathbb{A}$:} For the solar farm, its action is the bidding $p_t^\text{S}$ based on the plant availability. The BESS's actions include the bid power in the spot market $p_t^\text{Bat,SM}$, the power planned to draw from onsite solar curtailments $\hat{p}_t^\text{Bat,SC}$, and the charge/discharge of the BESS $v_t^\text{Ch}/v_t^\text{Dch}$. To stabilize the DRL training process, all power-related variables, e.g., $p_t^\text{S}$, $p_t^\text{Bat,SM}$, and $\hat{p}_t^\text{Bat,SC}$, are normalized into the range of $[0,1]$ and denoted as $a_t^\text{S}$, $a_t^\text{Bat,SM}$, and $a_t^\text{Bat,SC}$, respectively. Such normalization operations make action spaces inherently satisfy optimization constraints defined from \eqref{eq:cons_solar_power} to \eqref{eq:cons_BESS_SC_power}. 
The actions of the solar farm and the BESS are expressed as
\begin{equation}
    \bm{a}_t^\text{S} = \left[ a_t^\text{S} \right], ~ \bm{a}_t^\text{Bat} = \left[ v_t^\text{Ch},v_t^\text{Dch},a_t^\text{Bat,SM},a_t^\text{Bat,SC} \right].
\end{equation}

\textbf{Probability Space $\mathbb{P}$:} The probability space represents a probability set of transitioning from the current state to the next state after taking an action at the current time, which can be expressed as $\mathbb{P}\left(\bm{s}_{t+1}|\bm{s}_t,\bm{a}_t\right)$.

\textbf{Reward Space $\mathbb{R}$:} A reward mechanism is adopted in the MDP to assess the effectiveness of the state transition, denoted by $\mathbb{R}\leftarrow \mathbb{S}\times \mathbb{A}\times \mathbb{S}$. As the goal of the DRL is to maximize the expected cumulative rewards of the MDP, designing a suitable reward function plays a pivotal role in making better bidding decisions to optimize the MDP.

To mitigate the uncertainty of solar PV generation and update accurate dispatch targets for solar revenue maximization, the reward function for the solar farm can be formulated as
\begin{equation}
    \label{eq:reward_func_solar}
    r_t^\text{S} = -\lambda_t\left|a_t^\text{S}-\frac{p_t^\text{S,Act}}{p_t^\text{S,Ava}}\right|.
\end{equation}

To perform effective energy arbitrage, we introduce two charge/discharge indicators, denoted by $\mathbb{I}^\text{Ch}_t$ and $\mathbb{I}^\text{Dch}_t$. The appropriate opportunities for arbitrage operations can be indicated by $\mathbb{I}_t^\text{Ch} = \text{sgn}(\bar{\lambda}_t-\lambda_t)$ and $\mathbb{I}_t^\text{Dch} = \text{sgn}(\lambda_t-\bar{\lambda}_t)$, respectively, where $\text{sgn}(\cdot)$ is the sign function and $\bar{\lambda}_t$ is the exponential moving average of the spot price. The moving average price $\bar{\lambda}_t$ is expressed as
\begin{equation}
    \label{eq:Spot_price_moving_avg}
    \bar{\lambda}_t = \tau\bar{\lambda}_{t-1} + \left(1-\tau\right)\lambda_t,
\end{equation}
where $\tau \in (0,1)$ is a smoothing parameter. The proposed charge/discharge indicators encourage the BESS to purchase energy at relatively lower spot prices, i.e., $\lambda_t<\bar{\lambda}_t$, and sell energy at high prices, i.e., $\lambda_t>\bar{\lambda}_t$. If the BESS does not follow such an arbitrage guideline to bid in the spot market, the charge/discharge indicators will be negative, resulting in negative rewards. Thus, the BESS's reward function for energy arbitrage can be formulated as
\begin{equation}
    \label{eq:reward_func_BESS_arbitrage}
    r_t^\text{Bat,SM} = a_t^\text{Bat,SM} |\lambda_t - \bar{\lambda}_t| \left(\mathbb{I}_t^\text{Ch}v_t^\text{Ch} + \mathbb{I}_t^\text{Dch}v_t^\text{Dch}\right).
\end{equation}

Moreover, the BESS is incentivized to store otherwise curtailed solar power from the onsite solar farm. The reward function for managing solar curtailments is designed as
\begin{equation}
    \label{eq:reward_func_BESS_SC}
    r_t^\text{Bat,SC} = \beta\lambda_t \frac{p_t^\text{Bat,SC}}{P_\text{max}^\text{Bat}} \frac{f_{t-1}^\text{SC}}{L},
\end{equation}
where $\beta$ is the incentive factor for curtailment reduction.

Also, the battery degradation cost defined in \eqref{eq:battery_deg_cost} is transformed into the below reward function, formulated as
\begin{equation}
    \label{eq:reward_func_BESS_deg}
    r_t^\text{Bat,Deg} = d^\text{deg}_t\left|a_t^\text{Bat,SM}+\frac{p_t^\text{Bat,SC}}{P_\text{max}^\text{Bat}}\right|.
\end{equation}

Combining reward functions for energy arbitrage, solar curtailments, and battery degradation, the overall reward function for the BESS can be written as
\begin{equation}
    \label{eq:reward_func}
    r_t^\text{Bat} = r_t^\text{Bat,SM} + r_t^\text{Bat,SC} - r_t^\text{Bat,Deg}.
\end{equation}
optimizing the BESS's revenue and mitigating battery degradation defined in \eqref{eq:revenue_BESS} and \eqref{eq:battery_deg_cost}, respectively.

\subsection{Optimizing MDPs via AC-DRL-empowered DDPG}\label{subsec:method_ACDRL}
\subsubsection{DDPG Preliminaries} \label{subsubsec:method_ACDRL_preliminary}
We use one of the most representative DRL algorithms, namely DDPG~\cite{lillicrap2015_ddpg}, to learn an optimal bidding strategy denoted by $\pi(\bm{a}_t|\bm{s}_t)$ to maximize the expected return of the derived MDP. Note that we implement the same DDPG structure for both the solar farm and the BESS. The objective of the action strategy can be defined as
\begin{equation}
    \label{eq:DRL_obj}
    J_\pi = \mathbb{E}_{\bm{s}_t\sim \mathbb{P},\bm{a}_t\sim \pi(\bm{s}_t)} \left[R_t\right],
\end{equation}
where $R_t=\sum_{t'=t}^T \gamma^{t'-t}r_{t'}$ is the expected return from the $t$-th time frame with $\gamma$ denoted as the discounted factor. Besides the action strategy $\pi$, the DDPG designs a function, known as the critic function or the Q function, to gather the joint state-action pair as input and assess its effectiveness. The critic function can be formulated using the Bellman equation as
\begin{equation}
    \label{eq:critic_func}
    \begin{aligned}
    Q\left(\bm{s}_t,\bm{a}_t\right) &= \mathbb{E}_{\bm{s}_t\sim \mathbb{P},\bm{a}_t\sim\pi(\bm{s}_t)} \left[R_t|\bm{s}_t,\bm{a}_t\right],\\
    &=r_t + \gamma \mathbb{E}_{\bm{s}_{t+1}\sim\mathbb{P},\bm{a}_{t+1}\sim\pi}\left[Q\left(\bm{s}_{t+1},\bm{a}_{t+1}\right)\right].
    \end{aligned}
\end{equation}

The DDPG algorithm estimates the action strategy and critic function using an actor network $\pi_\psi(\bm{a}_t,\bm{s}_t)$ and a critic network $Q_\theta(\bm{s}_t,\bm{a}_t)$ parameterized with $\psi$ and $\theta$, respectively. The Adam optimizer is often utilized to update the parameters of neural networks via gradient descent, where the actor network is updated as $\psi \leftarrow \psi + \eta_\psi \nabla_\psi J_{\pi_\psi}$, where $\eta_\psi$ is the learning rate of the actor network. The critic network is updated by minimizing the residual error of the Bellman equation, where a target critic network (parameterized by $\hat{\theta}$)  is introduced to estimate the right side of \eqref{eq:critic_func}. The loss function of the critic network can be written as
\begin{equation}
    \label{eq:critic_net_loss_func}
    \hspace{-0.5mm}L(\theta) = \mathbb{E}\left\{\left\{Q_\theta\left(\bm{s}_t,\bm{a}_t\right)-\left[r_t+Q_{\hat{\theta}}\left(\bm{s}_{t+1},\bm{a}_{t+1}\right)\right]\right\}^2\right\}.
\end{equation}
Thus, the updating process of the critic network is defined as $\theta \leftarrow \theta - \eta_\theta \nabla_\theta L(\theta)$, where $\eta_\theta$ is the learning rate of the critic network. Moreover, the target network is updated using parameters of the critic network periodically in a moving average manner. The neural network adopted in DDPG is commonly a multi-layer perceptron (MLP), consisting of multiple fully-connected neural network layers (FCNNL).

\subsubsection{AC-DRL Network Structure} \label{subsubsec:method_ACDRL_ACDRL}
In our co-located solar-battery system, the bidding decisions of both the solar farm and the BESS are driven by multiple features as MDPs' states defined in \eqref{eq:MDP_state_solar} and \eqref{eq:MDP_state_BESS}. For instance, the time-varying spot price incentivizes the BESS to buy low and sell high for energy arbitrage. Meanwhile, its arbitrage behaviors may also be limited by solar curtailment management since the BESS needs to allocate the charging space for both market participation and onsite solar curtailments. By exploring the correlations among features and emphasizing their relative importance during the decision-making process, we can make more informed bidding decisions. This allows the BESS to pay greater attention to critical features inside the input MDP state, resulting in better bidding outcomes. 

\begin{figure}[!t]
    \centering
    \includegraphics[width=.85\linewidth]{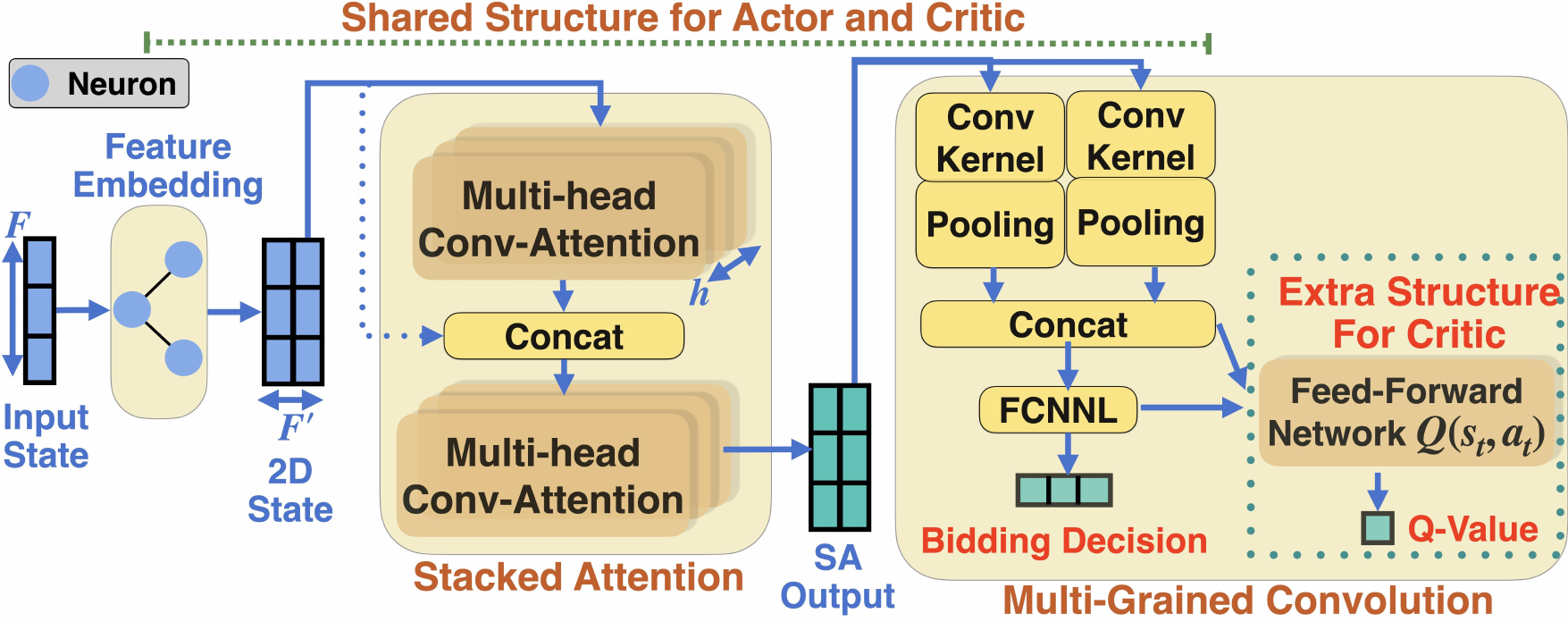}
    \caption{The AC-DRL framework.}
    \label{fig:framework_AC-DRL}
\end{figure}

However, the typical MLP-DRL structure fails to capture such \textit{correlation} and \textit{importance} information~\cite{vaswani2017}. We are thus motivated to develop a novel AC-DRL neural network structure for the DDPG algorithm, which consists of two main components: a stacked attention mechanism and a multi-grained feature convolution module, as depicted in Fig. \ref{fig:framework_AC-DRL}. Moreover, the actor and critic networks are unified to share most of neural network layers of the AC-DRL, reducing the number of trainable parameters and subsequently accelerating the model's training and inference speed. 

\begin{figure}[!t]
    \centering
    \includegraphics[width=.85\linewidth]{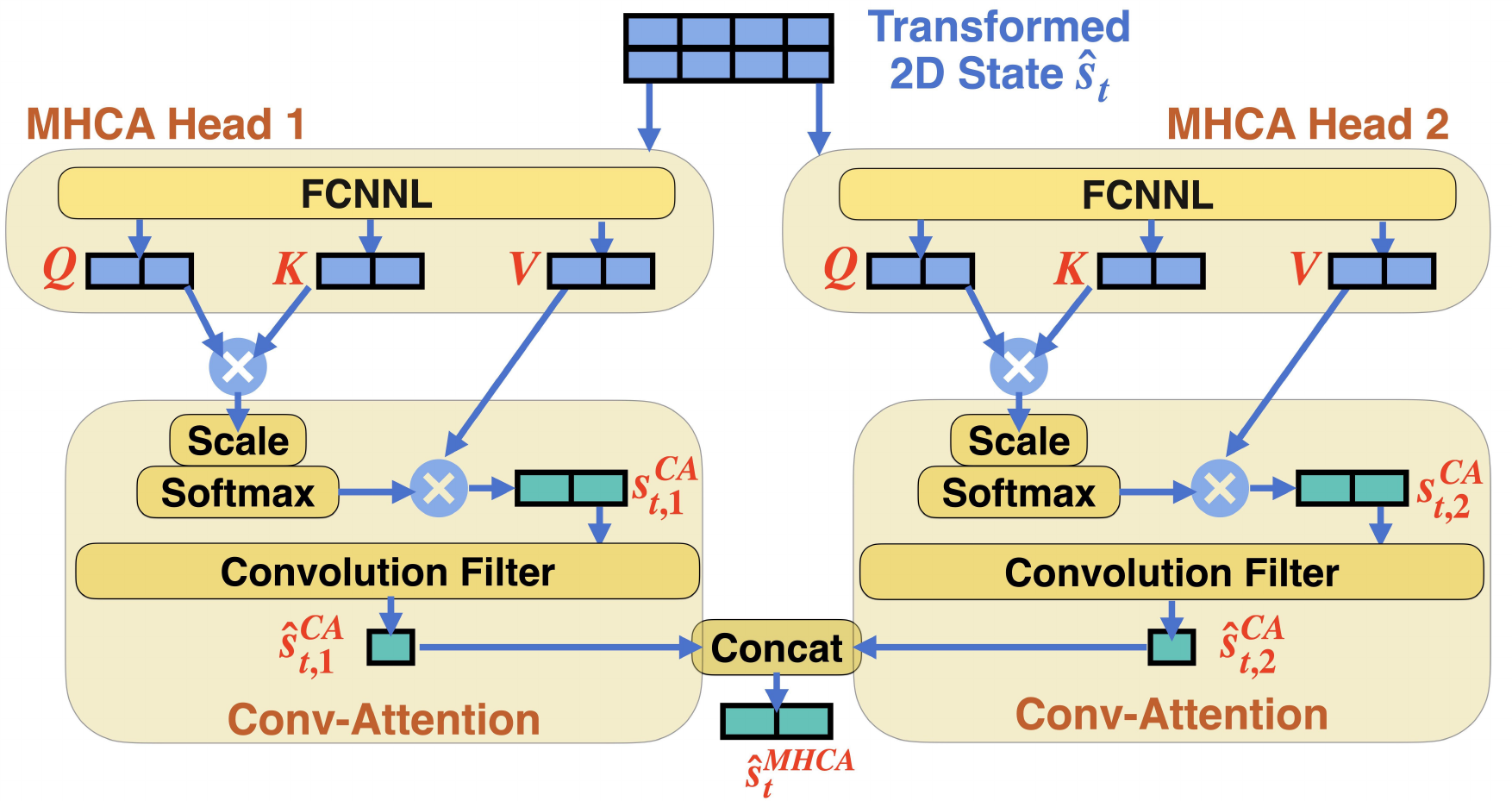}
    \caption{The inner structure of one MHCA with two heads.}
    \label{fig:two_head_MHCA}
\end{figure}

\textbf{Stacked Attention}: Before fed into the stacked attention, the input state $\bm{s}_t$ first passes through a \textit{feature embedding} layer as shown in Fig. \ref{fig:framework_AC-DRL}. The input state is projected into a two-dimensional feature space using one FCNNL, with each feature encoded as a specific one-dimensional embedding. The embedding layer can be formulated as $\hat{\bm{s}}_t=FCNNL(\bm{s}_t)\in\mathbb{R}^{F\times F'}$, where $F$ is the number of features, i.e., the length of input state, and $F'$ is the length of feature embedding. 

The stacked attention, which contains multiple multi-head convolutional attention (MHCA), takes the embedded state as input, explores the correlations between every two features, and calculates their mutual importance for bidding decisions. 

One MHCA divides the transformed feature space $F'$ into multiple sub-spaces $\frac{F'}{h}$, also known as the heads of the MHCA, where $h$ is the number of heads. We illustrate the inner structure of one MHCA with two heads in Fig. \ref{fig:two_head_MHCA}, where each head individually processes a low-dimensional feature map. The embedded state $\hat{\bm{s}}_t$ passes through all heads concurrently to create the query, key, and value matrices denoted by $Q$, $K$, and $V$, which prepares for correlation computation. Such a linear projection process can be defined as
\begin{align}
    \label{eq:query_mat}
    Q_{t,j} &= ReLU\left(W_{Q_j}\hat{\bm{s}}_t+b_{Q_j}\right) \in\mathbb{R}^{F\times \frac{F'}{h}},\\
    \label{eq:key_mat}
    K_{t,j} &= ReLU\left(W_{K_j}\hat{\bm{s}}_t+b_{K_j}\right) \in\mathbb{R}^{F\times \frac{F'}{h}},\\
    \label{eq:val_mat}
    V_{t,j} &= ReLU\left(W_{V_j}\hat{\bm{s}}_t+b_{V_j}\right) \in\mathbb{R}^{F\times \frac{F'}{h}},
\end{align}
where $j=1,\cdots,h$ is the index of MHCA head, $ReLU(x)=\max\{0,x\}$ is the activation function, $W_{Q_j}$, $W_{K_j}$, and $W_{V_j}$ are learnable weighted matrices, $b_{Q_j}$, $b_{K_j}$, and $b_{V_j}$ are learnable bias matrices. The generated matrices are then employed by the \textit{convolutional attention} to calculate the correlation strength of every two feature embeddings, which can be formulated as 
\begin{equation}
    \label{eq:self_attn}
    \bm{s}_{t,j}^\text{CA} = softmax\left(\frac{Q_{t,j}K_{t,j}^T}{\sqrt{F'}}\right)V_{t,j} = W_{t,j}^\text{att}V_{t,j},
\end{equation}
where the result of the softmax operation, denoted by $W_{t,j}^\text{att}\in\mathbb{R}^{F\times F}$, is known as the attention weight matrix, whose element represents the mutual influences of a feature pair. The multiplication between the attention matrix and the value matrix aims to aggregate such correlation information to its original input state $\hat{\bm{s}}_t$ and indicate the relative importance of each feature. Unlike the original self-attention mechanism~\cite{vaswani2017}, we further deploy a convolution layer, denoted by $Conv(x)$ to directly extract the aforementioned correlation information. Concatenating the outputs of each head, we obtain the final output of one MHCA as 
\begin{equation}
    \label{eq:MHCA}
    \bm{s}_{t,i}^\text{MHCA} = Concat\left(Conv\left(\bm{s}_{t,1}^\text{CA}\right),\cdots,Conv\left(\bm{s}_{t,h}^\text{CA}\right)\right),
\end{equation}
where $i=1,\cdots,N_\text{MHCA}$ is the index of MHCA and $N_\text{MHCA}$ is the number of MHCA in our stacked attention mechanism. Moreover, the original input information will inevitably lose within the stacked attention since the convolutional operation downsamples the input data. To keep the MHCA remembering the full information, we also concatenate the transformed two-dimensional input $\bm{s}_t$ with each MHCA's output $\bm{s}_{t,i}^\text{MHCA}$ (except for the last MHCA), which works as the input of the following MHCA, as shown in the dotted blue line in Fig. \ref{fig:framework_AC-DRL}.

\textbf{Multi-grained Feature Convolution}: We denote the output of the stacked attention as $\bm{s}_{t,N_\text{MHCA}}^\text{MHCA}$. Taking advantage of the ``attention'' information, we develop multi-grained convolution filters, as shown in Fig. \ref{fig:framework_AC-DRL}, to fully exploit and extract the hidden information of the two-dimensional attention output for making better bidding decisions. Each convolution filter is connected with a max-pooling layer to downsample and aggregate the extracted features into one-dimensional space. The overall feature convolution process can be formulated as 
\begin{equation}
    \label{eq:conv}
    \bm{s}_{t,k}^\text{Conv} = MaxPooling\left(\bm{h}_{\phi_k} * \bm{s}_{t,N_\text{MHCA}}^\text{MHCA}\right),
\end{equation}
where $*$ represents the convolution operator and $\bm{h}_{\phi_k}$ is a two-dimensional convolution filter parameterized with $\phi_k$, where $k=1,\cdots,N_\text{Conv}$ is the index of the convolution filter with $N_\text{Conv}$ denoting the number of multi-grained convolution filters. As the size of a convolution filter is determined by its parameters, a group of filter parameters, i.e., $\{\phi_1,\cdots,\phi_k\}$, enables our convolution operation to gather various lower-scale feature maps and subsequently extract multi-grained feature information from the underlying attention output.

We then aggregate the extracted feature maps from each convolution filter and generate our bidding decisions through one more FCNNL, which can be formulated as
\begin{equation}
    \label{eq:bidding_decision_output}
    \bm{a}_t = FCNNL\left(Concat\left(\left[\bm{s}_{t,1}^\text{Conv},\cdots,\bm{s}_{t,N_\text{Conv}}^\text{Conv}\right]\right)\right).
\end{equation}

For the critic network $Q_\theta$, the bidding decisions are integrated with multi-grained convolution results, passing through a \textit{Feed-Forward} network (comprising two FCNNLs activated by the ReLU function) to produce a Q value, as illustrated in Fig. \ref{fig:framework_AC-DRL}.

\begin{table}[!t]
    \centering
    \caption{Initialized parameters.}
    \begin{tabular}{cc||cc||cc}
    
    \hline
    
    $\Delta t$ & $5$ mins & $\alpha$ & $1.5$ & $c$ & $1$ AU\$/MWh \\ 
    
    $P_\text{max}^\text{Bat}$ & $10$ MW & $P_\text{max}^\text{S}$ & $65$ MW & $\sigma$ & $0.625$ \\
    
    $E_\text{min}^\text{Bat}$ & $0.5$ MWh & $E_{1,\text{max}}^\text{Bat}$ & $9.5$ MWh & $\eta^\text{Ch},\eta^\text{Dch}$ & $0.95$ \\ 
    
    $H$ & $2016$ & $\tau$ & $0.9$ & $\beta$ & $6$ \\
    
    $L$ & $10$ & $\gamma$ & $0.99$ & $F'$ & $64$ \\ 
    
    $h$ & $8$ & $N_\text{MHCA}$ & $2$  & $N_\text{Conv}$ & $5$ \\ 
    
    \hline
    
    \end{tabular}
    \label{tab:parameters}
\end{table}

\section{Experiments and Results} \label{sec:exps}
\subsection{Experimental Setup} \label{subsec:exps_setup}

We use solar PV generation data in a five-minute resolution from the Rugby Run Solar Farm located in Queensland, Australia in 2020. We also use the spot price of the Queensland~\cite{aemo_dashboard}, which is the jurisdiction of the solar farm, to ensure that all data are well aligned when training and testing our AC-DRL method and the MLP-DRL benchmark. In particular, the first eleven months of data in 2020 are used for training and the last month is for evaluation. 
The storage capacity of the BESS is set to be $10$ MWh with its minimum and maximum allowable state of charge being $5\%$ and $95\%$, respectively. The period length $H$ of updating the battery degradation coefficient is set to be one week. 
We used an Nvidia TITAN RTX graphics processing unit for algorithm training. A Gaussian noise following the distribution $\mathcal{N}\sim(0,0.1)$ is adopted in the DDPG algorithm. For the hyperparameters of our DRL training, the batch size is $512$. The learning rates for both actor and critic networks are $8\text{e}^{-4}$. The replay buffer size is $1\text{e}^5$. 
Moreover, we rename our AC-DRL strategy as AC-DDPG in this section (so as well for the MLP-DRL method into MLP-DDPG), as we use other representative DRL algorithms for comparisons in Section \ref{subsec:exps_benchmark_drlbaseline_comp}. The initialized parameters are provided in Table \ref{tab:parameters}.

\begin{table}[!t]
    \centering
    \caption{The evaluation revenues (in AU\$) of the DMPC, SMPC, MLP-DDPG, and our AC-DDPG methods.}
    \begin{tabular}{ c c c c}
     & Solar & BESS & Solar\&BESS \\
     
     \hline
     
     DMPC & $533,302$ & $62,277$ & $595,579$ \\
     
     \hline
     
     SMPC & $548,391$ & $64,156$ & $612,547$ \\
     
     \hline 
     
     MLP-DDPG & $586,413$ & $74,862$ & $661,275$ \\
     
     \hline
     
     \textbf{AC-DDPG (Ours)} & $\bm{638,934}$ & $\bm{95,881}$ & $\bm{734,815}$\\
     
     \hline
     
    \end{tabular}
    \label{tab:benchmark_rev}
\end{table}

\begin{figure}[!t]
    \centering
    \includegraphics[width=.8\linewidth]{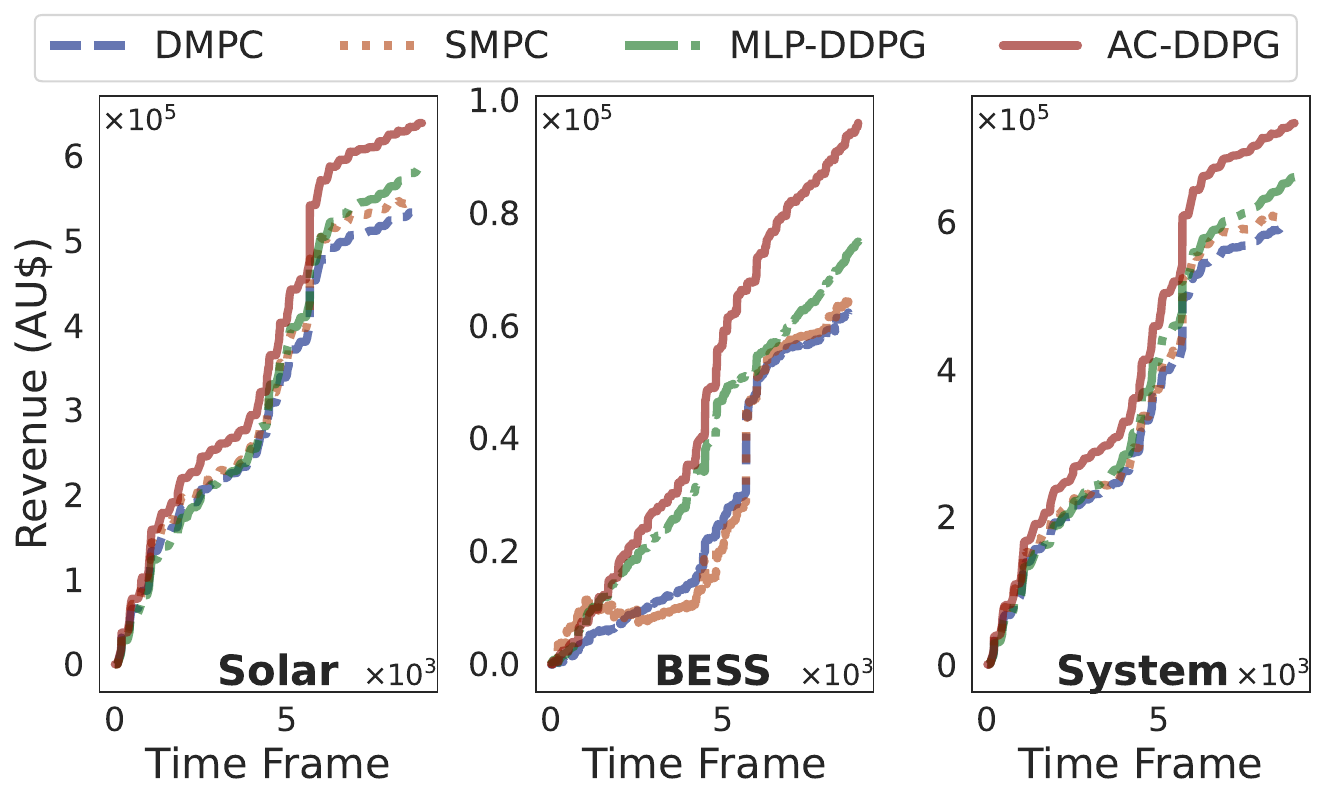}
    \caption{Evaluation revenue comparisons of the DMPC benchmark, the SMPC benchmark, the MLP-DDPG benchmark, and our AC-DDPG.}
    \label{fig:benchmark_rev}
\end{figure}

\subsection{Comparisons with MPC Benchmarks and DRL Baselines} \label{subsec:exps_benchmark_drlbaseline_comp}

\begin{figure}[!t]
    \centering
    \includegraphics[width=.8\linewidth]{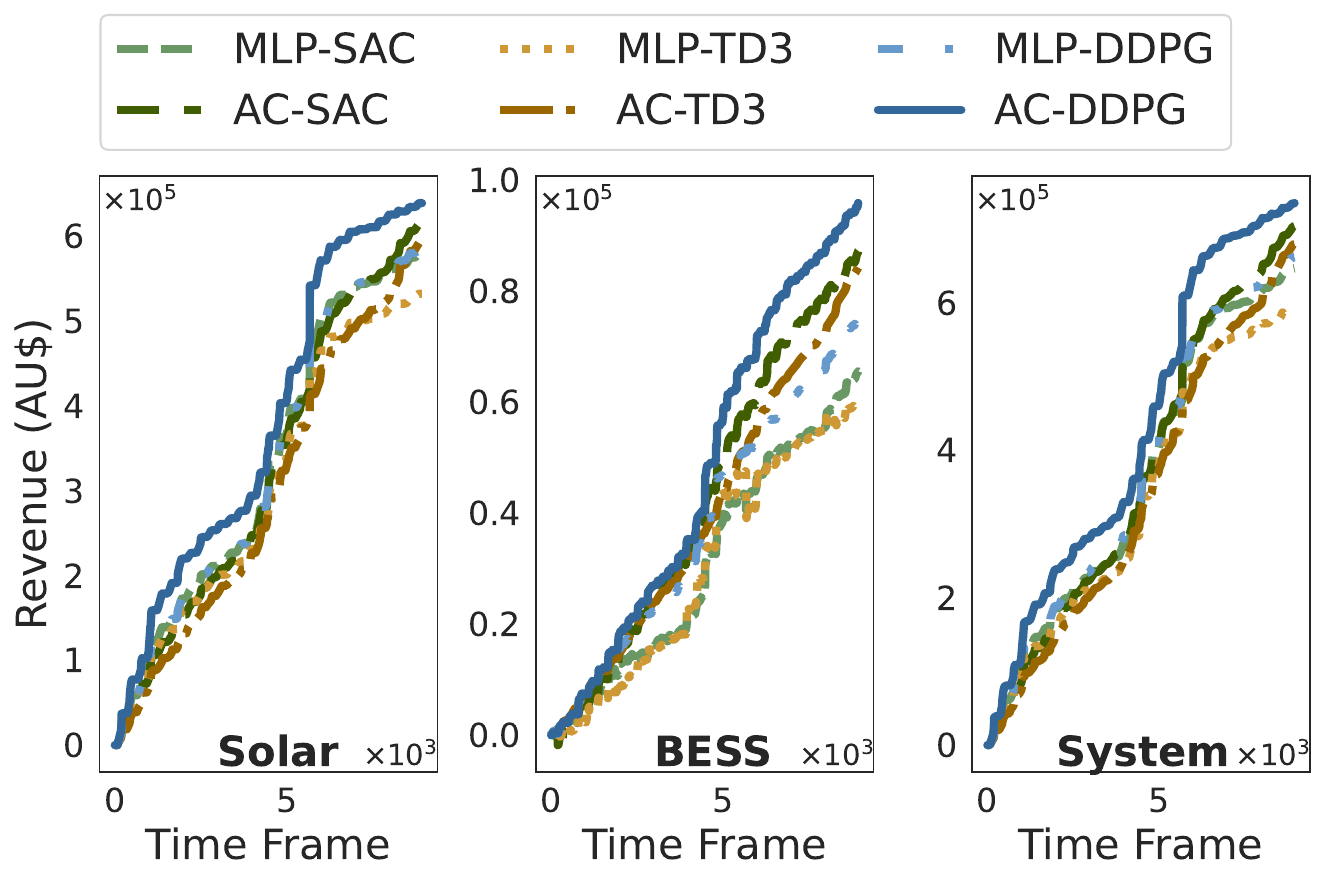}
    \caption{Evaluation revenue comparisons of DDPG, SAC, and TD3 algorithms with/without the AC mechanism.}
    \label{fig:drl_baseline_rev}
\end{figure}

To evaluate the effectiveness of the proposed AC-DDPG, besides the MLP-DDPG benchmark, we develop the DMPC benchmark~\cite{badoual2021_DMPC}, which is a typical MPC-based approach relying on deterministic forecasts. Specifically, we use a gated recurrent unit (GRU) network to produce forecast results of solar PV generation and the spot price in the next $50$ dispatch intervals, based on a one-day length of historical input. The predictions are then fed into a mixed integer linear programming solver (empowered by the Gurobi solver from the PuLP library~\cite{mitchell2011}) to solve the bidding problem defined from \eqref{eq:optimization_obj} to \eqref{eq:cons_BESS_energy_change} for the solar-battery system. As the DMPC benchmark fails to tackle the uncertainties of solar generation and spot price, we also develop another MPC-based benchmark, namely the stochastic MPC (SMPC)~\cite{conte2018_SMPC}, to capture such uncertainties via scenario generation. The number of scenarios is $120$. 
The evaluation revenue comparisons of the benchmarks (DMPC, SMPC, and MLP-DDPG) and our AC-DDPG are depicted in Fig.~\ref{fig:benchmark_rev}, with the associated results also presented in Table~\ref{tab:benchmark_rev}. The evaluation outcomes show that the AC-DDPG significantly outperforms all three benchmarks, with substantial revenue boosts by $20\%$ for the solar farm, $54\%$ for the BESS, and $23\%$ for the whole system, compared to the DMPC method, by $17\%$, $49\%$, $20\%$ compared to the SMPC method, and by $9\%$, $28\%$, $11\%$ compared to the MLP-DDPG method, respectively. Moreover, we also compare the performance of the adopted DDPG algorithm with two representative off-policy algorithms -- the twined delayed deep deterministic policy gradient (TD3)~\cite{fujimoto2018_td3} and the soft actor critic (SAC)~\cite{haarnoja2018_sac}, which are trained with/without the attentive convolutional (AC) mechanism. The results are shown in Fig.~\ref{fig:drl_baseline_rev}, 
where the DDPG algorithm achieves higher bidding outcomes than the SAC and TD3 with/without the AC mechanism. Such simulations also demonstrate the effectiveness of our AC mechanism in improving the bidding performance of the co-located system from two aspects: 1) there are considerable revenue boosts with the introduction of the AC mechanism for all DRL algorithms, compared to the MLP-based neural network structure; 2) the performance of these AC-DRL approaches surpass the MPC-based benchmarks, i.e., DMPC and SMPC, by significant margins, suggesting the consistency of AC-DRL framework in achieving higher financial returns for the solar-battery system in electricity market participation.

\begin{figure}[!t]
    \centering
    \subfloat[No. Of Curtail/Response.]{
    \includegraphics[width=0.43\linewidth]{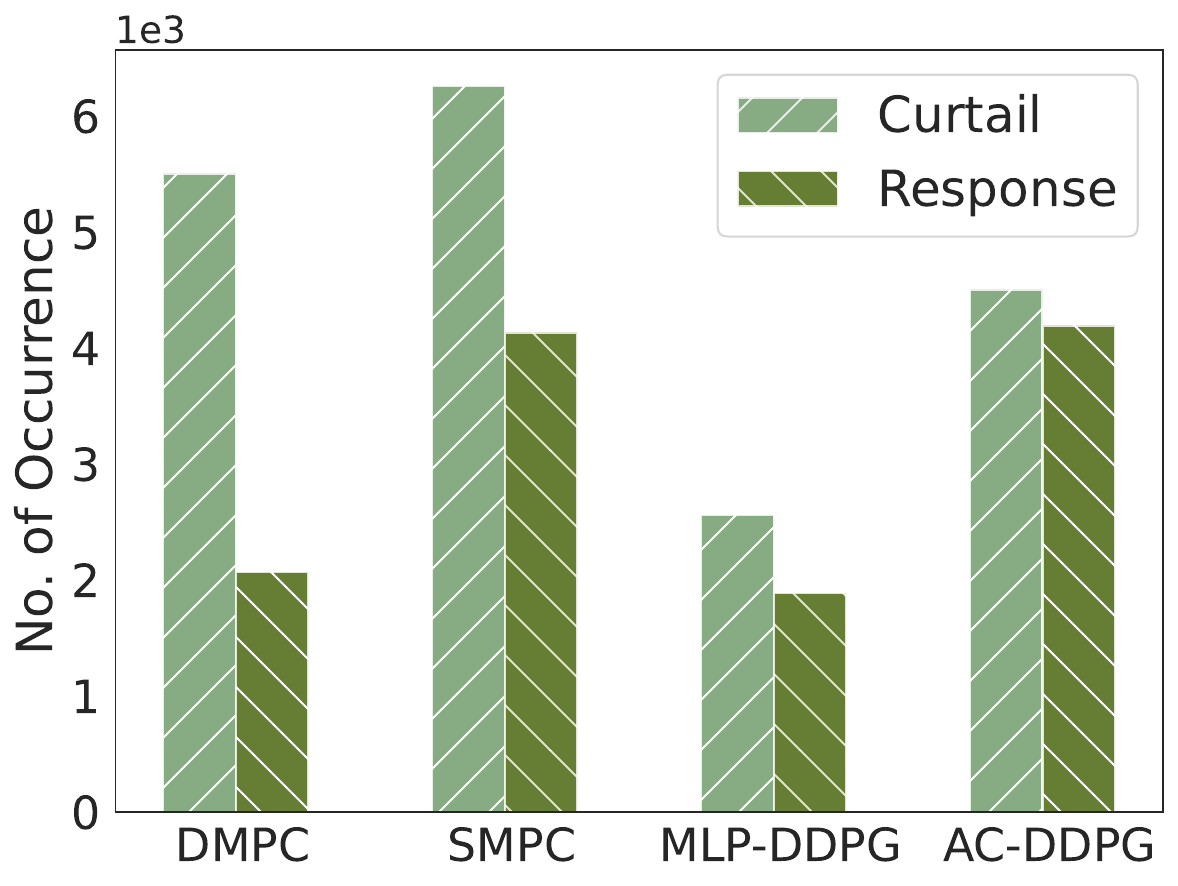}
    \label{fig:curtail_response_times}
    }
    \subfloat[Curtailed/Absorbed Energy.]{
    \includegraphics[width=0.43\linewidth]{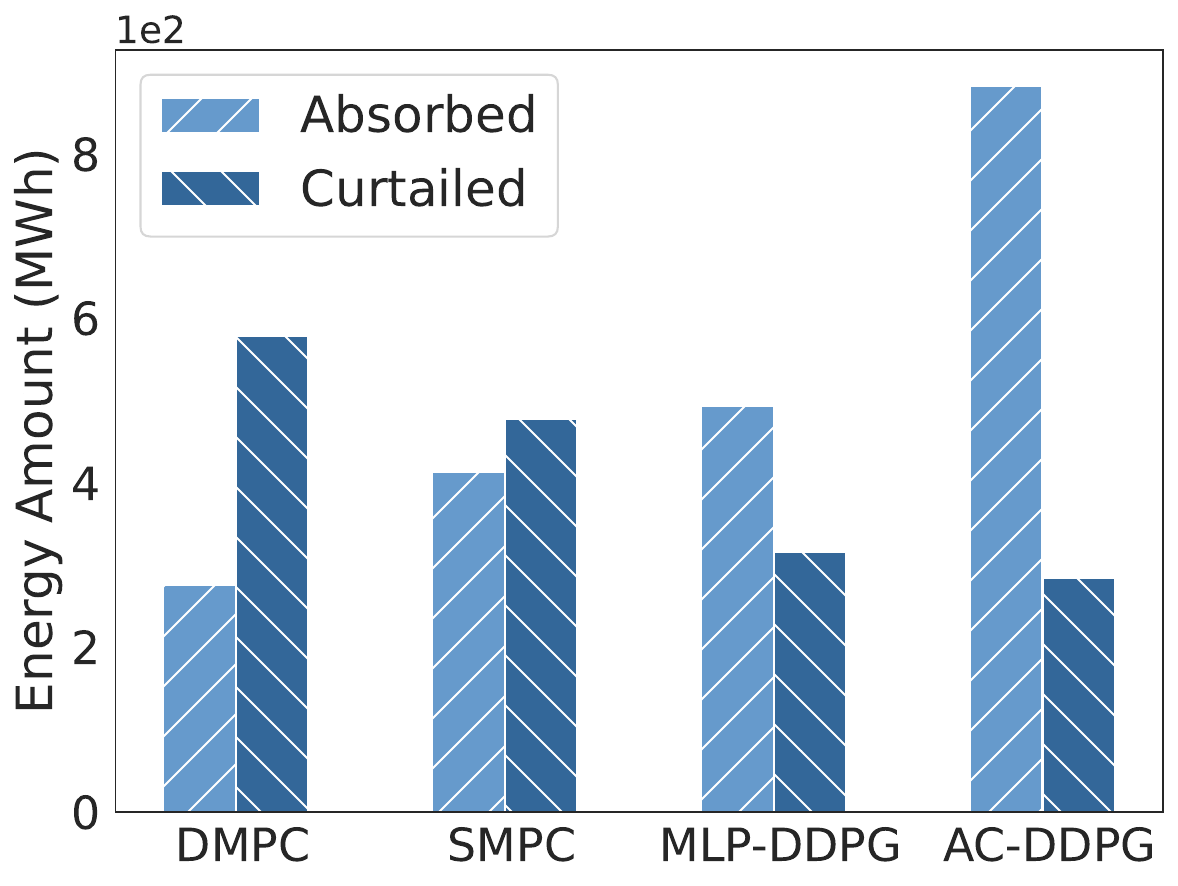}
    \label{fig:curtail_response_amount}
    }
    \vspace{-1.5mm}
    \caption{Solar curtailment management comparisons of the DMPC, SMPC, MLP-DDPG, and our AC-DDPG methods.}
    \label{fig:curtailment_management_comparisons}
\end{figure}

Besides revenue comparisons, to further analyze solar curtailment management of our AC-DDPG, we summarize key metrics related to curtailment management. These metrics include the number of solar curtailment events, the number of BESS responses to solar curtailment events (i.e., the number of times that the BESS successfully absorbs the extra solar generation that will be otherwise curtailed), the amount of absorbed curtailed solar energy by the BESS (in MWh), the amount of curtailed solar energy of the solar farm (in MWh), as shown in Fig. \ref{fig:curtailment_management_comparisons} and Table \ref{tab:curtailment_management_comparisons}. Note that all these statistics are calculated based on the whole evaluation process. The results in Fig. \ref{fig:curtail_response_times} indicate that our AC-DDPG is more responsive to solar curtailments. It successfully responds to approximately $93\%$ of solar curtailment events, which is notably higher than the DMPC, SMPC, and MLP-DDPG benchmarks by $55\%$, $27\%$, and $19\%$, respectively. Thus, the heightened responses to solar curtailments lead to a substantial increase in the amount of absorbed onsite curtailed solar energy, as shown in Fig. \ref{fig:curtail_response_amount}. The ``Curtailed'' bar represents the amount of otherwise curtailed solar energy after solar curtailment reduction, where all three benchmarks curtail more solar PV generation than our AC-DDPG method. Therefore, the curtailment management results demonstrate the significant capability of the proposed AC-DDPG in reducing solar curtailments and effectively harnessing onsite curtailed solar energy to enhance the economic performance of the co-located solar-battery system.

\subsection{The BESS Bidding Behavior Analysis} \label{subsec:exps_BESS_bidding_behavior}
In this section, we analyze how the BESS balances the trade-off between energy arbitrage in Section \ref{subsubsec:exps_BESS_bidding_behavior_arbitrage} and solar curtailment management in Section \ref{subsubsec:exps_BESS_bidding_behavior_curtailment}.

\begin{table}[!t]
    \centering
    \caption{Solar curtailment management comparisons of the DMPC, SMPC, MLP-DDPG, and our AC-DDPG methods.}
    \begin{tabular}{ c c c c c}
     \hline
     
     & \makecell{No. Of \\ Curtail} & \makecell{No. Of \\ Response} & \makecell{Absorbed \\ Energy}  & \makecell{Curtailed \\ Energy } \\
     
     \hline

     DMPC & $5498$ & $2066$ & $276$ MWh  & $579$ MWh \\

     \hline

     SMPC & $6256$ & $4129$ &  $413$ MWh  & $478$ MWh \\

     \hline

     MLP-DDPG & $2558$ & $1886$ & $493$ MWh  & $316$ MWh \\

     \hline

     \textbf{AC-DDPG (Ours)} & $4502$ & $4192$ & $\bm{883}$ MWh  & $\bm{285}$ MWh \\

     \hline
    \end{tabular}
    \label{tab:curtailment_management_comparisons}
\end{table}

\subsubsection{BESS Energy Arbitrage} \label{subsubsec:exps_BESS_bidding_behavior_arbitrage}
\begin{figure}[!t]
    \centering
    \includegraphics[width=.85\linewidth]{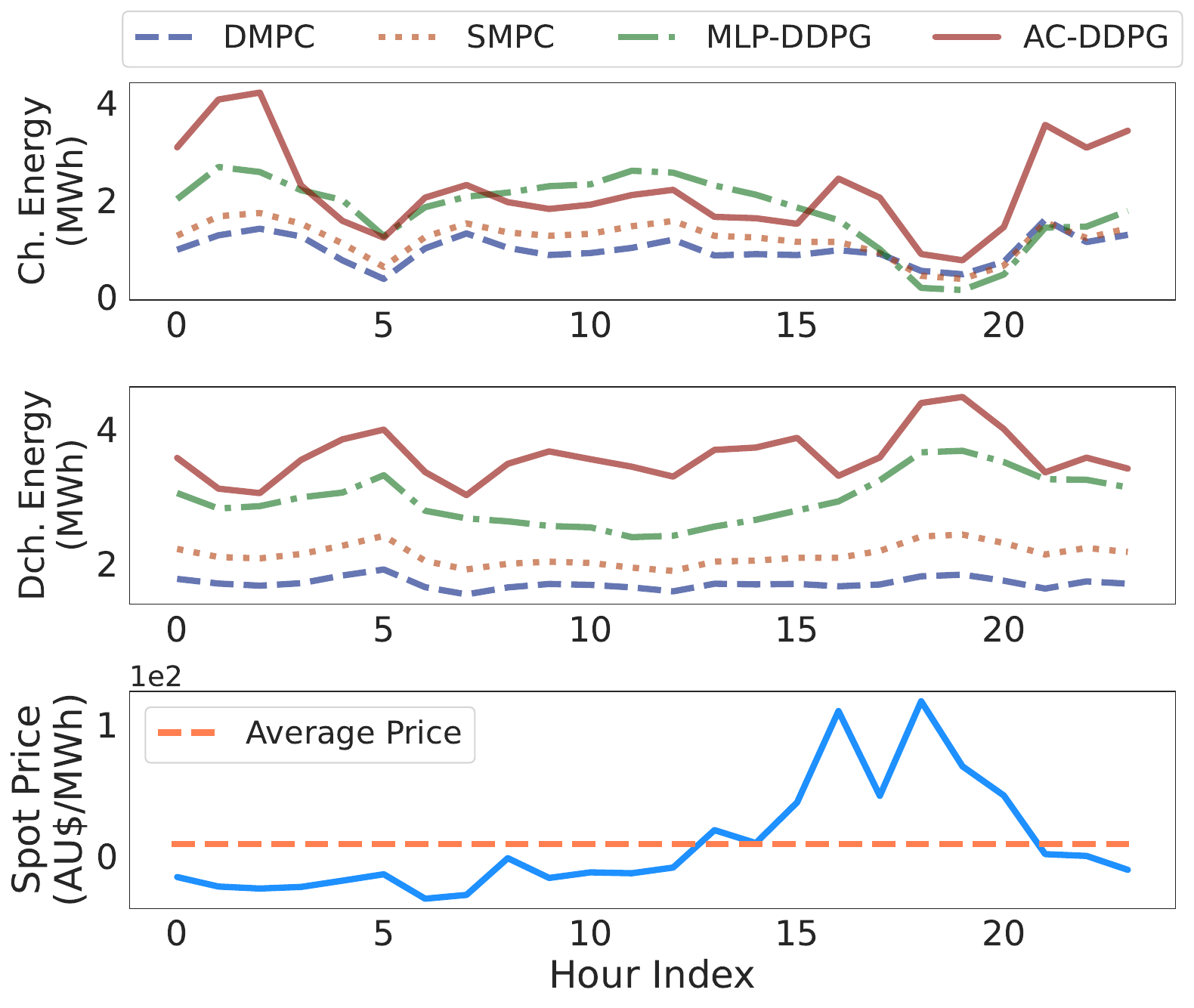}
    \caption{BESS's average hourly energy exchange with spot market.}
    \label{fig:BESS_hourly_dch_ch_bidding}
\end{figure}

The BESS performs energy arbitrage by buying energy at low prices (i.e., charge) and selling stored energy at high prices (i.e., discharge) in the spot market. The average hourly energy exchange with the spot market via charging/discharging operations throughout the evaluation process is depicted in Fig. \ref{fig:BESS_hourly_dch_ch_bidding}, along with the average hourly spot price. 
Charged/discharged energy at each hour index is accumulated via all $12$-interval operations in this hour. 

\begin{figure}[!t]
    \centering
    \includegraphics[width=.85\linewidth]{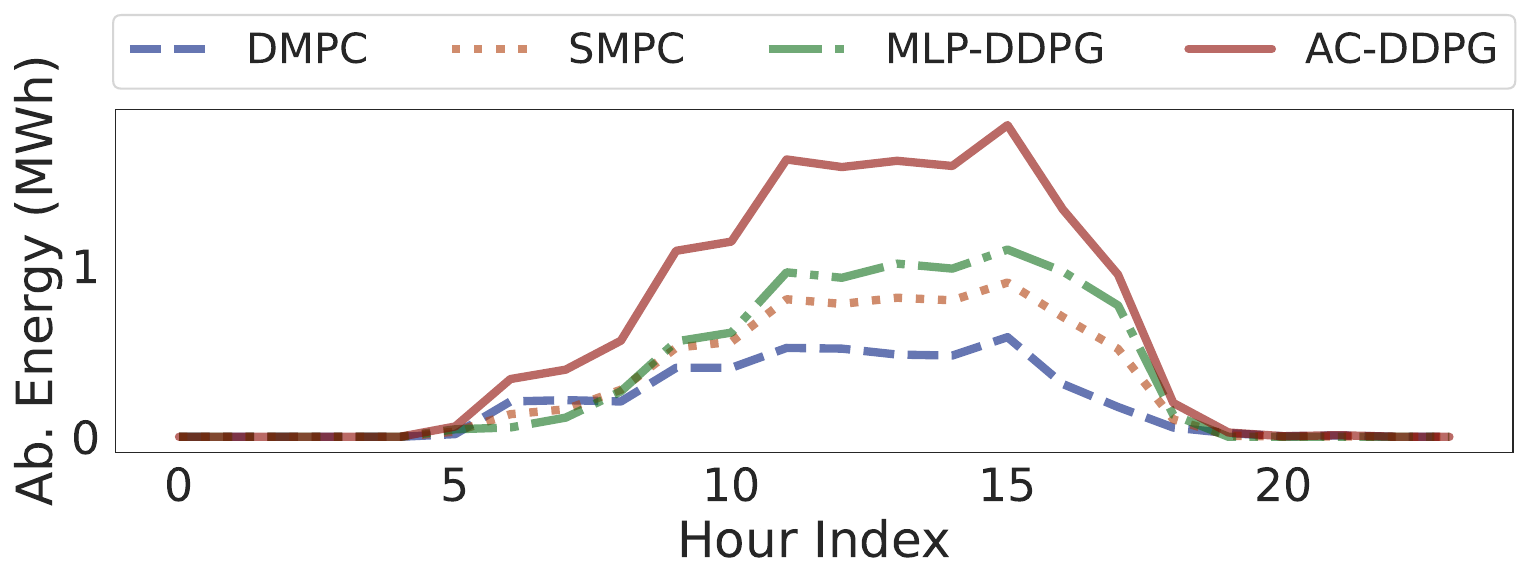}
    \caption{BESS's average hourly absorbed solar energy.}
    \label{fig:BESS_hourly_absorbed_curtailed_energy}
\end{figure}

The results show that, while the DMPC and SMPC methods appear to deviate from the buy-low-sell-high principle for arbitrage, both MLP-DDPG and AC-DDPG tend to purchase more energy during the daytime when the spot price stays at a relatively low level and subsequently sell energy during night peak hours, such as 7 p.m., when the spot price experiences a rapid increase. The BESS's strategy of the buy-low-sell-high arbitrage is also corroborated by the higher economic returns presented in Table \ref{tab:benchmark_rev}. Furthermore, our AC-DDPG method enables the BESS to conduct more frequent arbitrage during the daytime compared to the MLP-DDPG benchmark, as depicted in Fig. \ref{fig:BESS_hourly_dch_ch_bidding}. This increased frequency is be attributed to AC-DDPG's significant curtailment management capability. We illustrate the average hourly absorbed energy (in MWh) of the BESS from solar curtailments in Fig.~\ref{fig:BESS_hourly_absorbed_curtailed_energy}, where our AC-DDPG enables the BESS to more effectively take advantage of otherwise curtailed solar energy. The absorption of curtailed solar energy effectively charges the BESS (sometimes without buying energy from the spot market) and ensures a sufficient energy storage level for arbitrage operations.

\subsubsection{Solar Curtailment Management} \label{subsubsec:exps_BESS_bidding_behavior_curtailment}
\begin{figure}[!t]
    \centering
    \includegraphics[width=.85\linewidth]{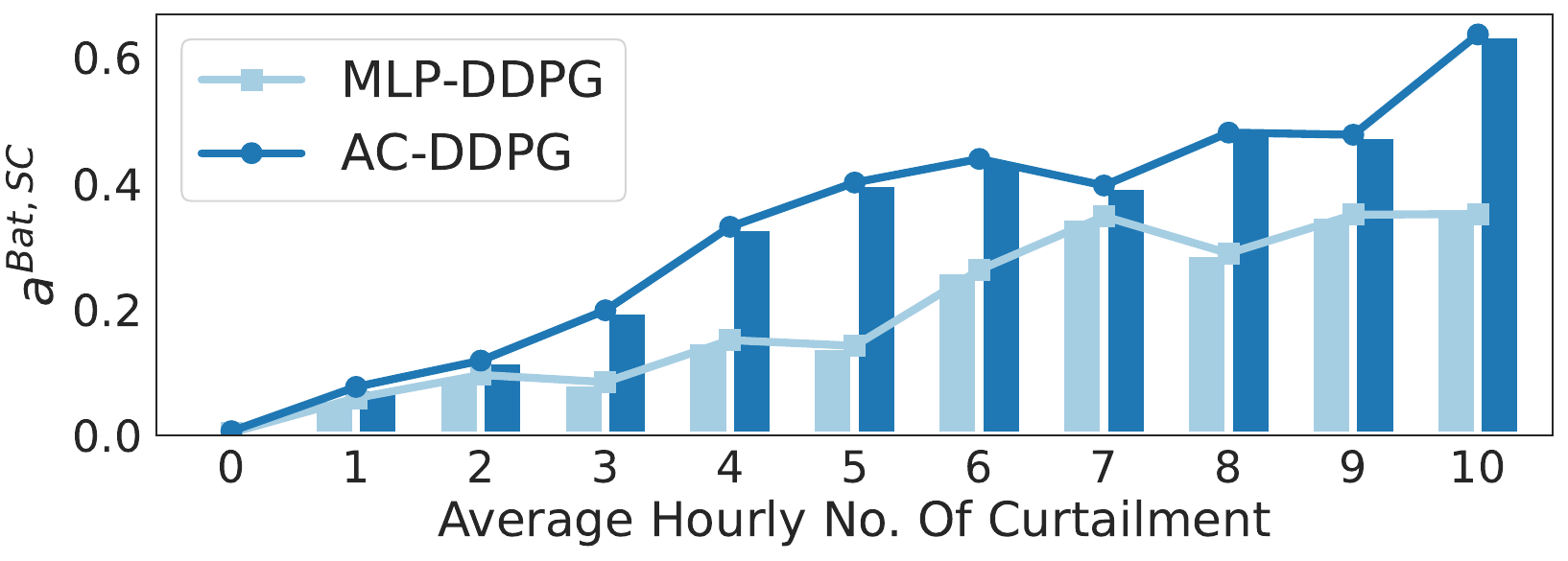}
    \caption{The solar curtailment management of MLP-DDPG and AC-DDPG under time-varying number of curtailment events.}
    \label{fig:BESS_hourly_reserve_curtail}
\end{figure}

We include the number of solar curtailment events (if there is no BESS) within the latest $10$ intervals in the BESS's state, as defined in \eqref{eq:MDP_state_BESS}, aiming to inform the BESS about the frequency of solar curtailments. By considering the time-varying number of curtailment events, the BESS can dynamically allocate the ratios between market bid $a_t^\text{Bat,SM}$ and power planned to be drawn from onsite solar curtailment $a_t^\text{Bat,SC}$,  striking a balance between energy arbitrage and curtailment management. The relationship between the average hourly number of curtailment events and the average allocated charging space for absorbing curtailment $a_t^\text{Bat,SC}$ during the evaluation process is depicted in Fig. \ref{fig:BESS_hourly_reserve_curtail}.  The results demonstrate that both MLP-DDPG and AC-DDPG enable the BESS to adjust the charging space allocated for solar curtailment reduction based on the recent occurrences of curtailments. As the number of curtailment events decreases, both methods allocate smaller charging space, while an increase in curtailment occurrences leads to a larger allocation. In contrast to MLP-DDPG, our AC-DDPG tends to be more proactive, allocating more charging space to account for potential onsite solar curtailments.

\begin{figure}[!t]
    \centering
    \includegraphics[width=.85\linewidth]{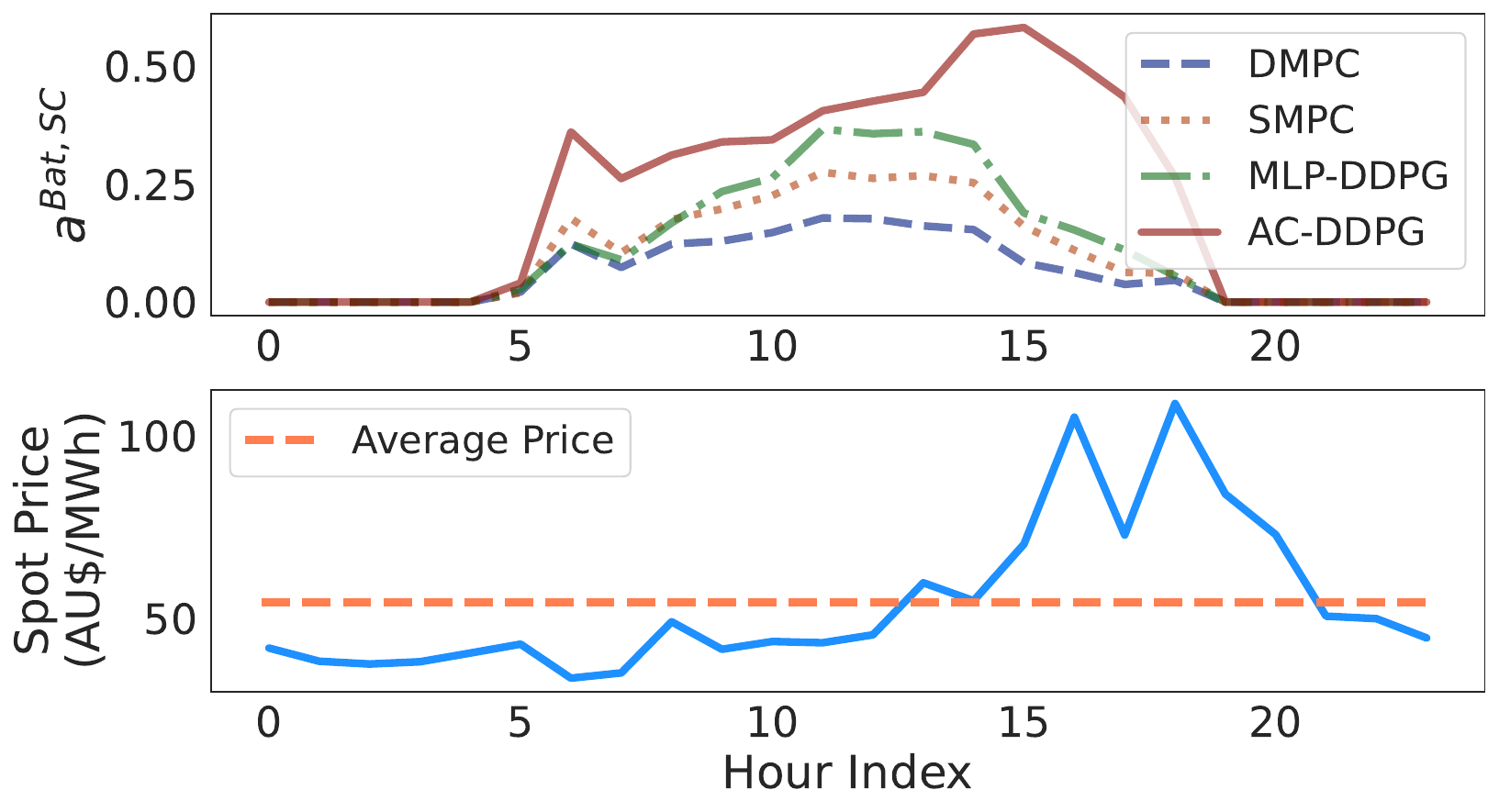}
    \caption{The solar curtailment management of DMPC, SMPC, MLP-DDPG, and AC-DDPG under time-varying spot prices.}
    \label{fig:BESS_hourly_reserve_price}
\end{figure}

BESS's solar curtailment management is also heavily influenced by the time-varying spot price. The reason is twofold. Firstly, energy arbitrage is the most direct revenue stream for the BESS and the arbitrage operation is sensitive to the spot price changes, as illustrated in Fig. \ref{fig:BESS_hourly_dch_ch_bidding}. Secondly, while the BESS allocates partial charging space to absorb the onsite curtailed solar generation instead of solely purchasing power from the spot market, solar curtailments may not occur due to the inherent variability of solar generation. Thus, coordinating energy arbitrage and curtailment reduction under highly stochastic spot prices is essential for maximizing profitability. We illustrate the impact of the average hourly spot price on curtailment management throughout the evaluation process in Fig. \ref{fig:BESS_hourly_reserve_price}. For the three benchmarks, as the spot prices increase (e.g., from $3$ p.m. to $8$ p.m.), the BESS tends to allocate less charging space for curtailment management, i.e., lower $a_t^\text{Bat,SC}$. In contrast, our AC-DDPG allocates more charging space under the same conditions. This indicates that our AC-DDPG favors curtailed solar energy during periods of higher spot price, as charging at higher spot prices is likely to result in significant economic losses. Using curtailed energy, though subject to availability, is free. Additionally, the BESS allocates sufficient charging space for the onsite solar curtailments, when the spot price is low and solar curtailments are likely to happen, such as from $10$ a.m. to $12$ a.m.

\section{Conclusion} \label{sec:conclusion}
This paper developed an effective coordination strategy for the co-located solar-battery system to enhance the system's economic performance in the spot market. We proposed a model-free DRL-based bidding strategy, namely AC-DRL, for the solar farm and the BESS to dynamically balance the trade-off between energy arbitrage and solar curtailment reduction, leveraging an attention mechanism and multi-grained feature convolution for feature correlation extraction. We validated the proposed AC-DRL using realistic solar farm data in the Australian NEM. Simulation results demonstrate that our method outperforms both the MPC-based and the DRL-based benchmarks in terms of achieving higher economic returns and improved management of solar curtailment. 

Our work also reveals several noteworthy observations: 1) The BESS strategically purchases more power to charge itself during daytime and sells power at night peak hours to take advantage of price spreads, adhering to the buy-low-sell-high arbitrage principle; {2) The BESS tends to allocate more charging space for potential solar curtailments in the presence of higher likelihood of solar curtailment events occurring; 3) the BESS tends to use more curtailed solar energy when the spot price rapidly increases; 4) the time-varying spot price is the main driver for the BESS's bidding decision-making.

\bibliographystyle{ieeetr}
\bibliography{IEEEabrv}

\begin{IEEEbiography}[{\includegraphics[width=1in,height=1in,clip,keepaspectratio]{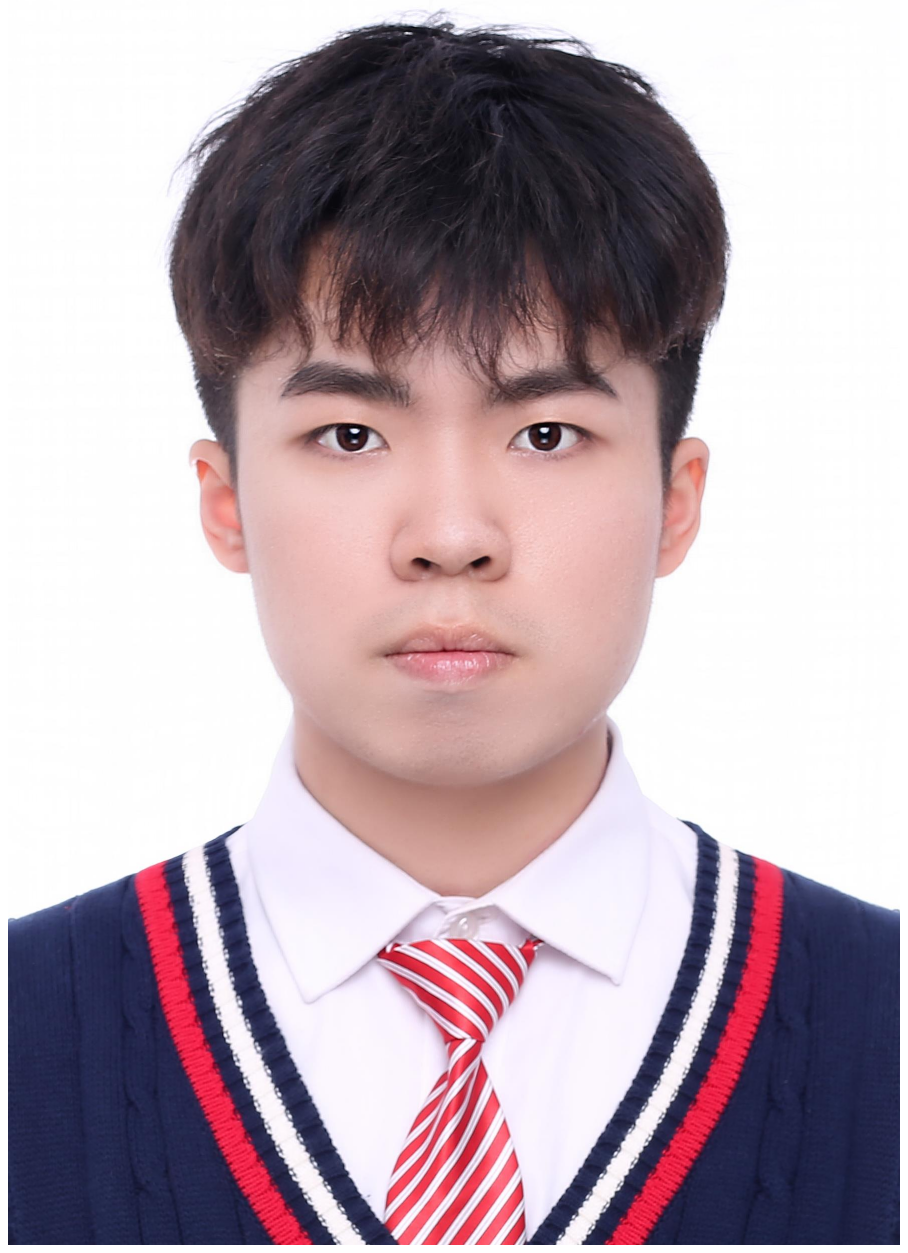}}]{Jinhao Li} received a B.E. degree in smart grid information engineering with a double B.S. degree in mathematics from the University of Electronic Science and Technology of China (UESTC) in 2022. He is currently pursuing his Ph.D. degree in Information Technology at the Department of Data Science and AI, Faculty of Information Technology, Monash University. His research interest is in machine learning for energy systems.
\end{IEEEbiography}

\vskip -3\baselineskip plus -1fil

\begin{IEEEbiography}[{\includegraphics[width=1in,height=1in,clip,keepaspectratio]{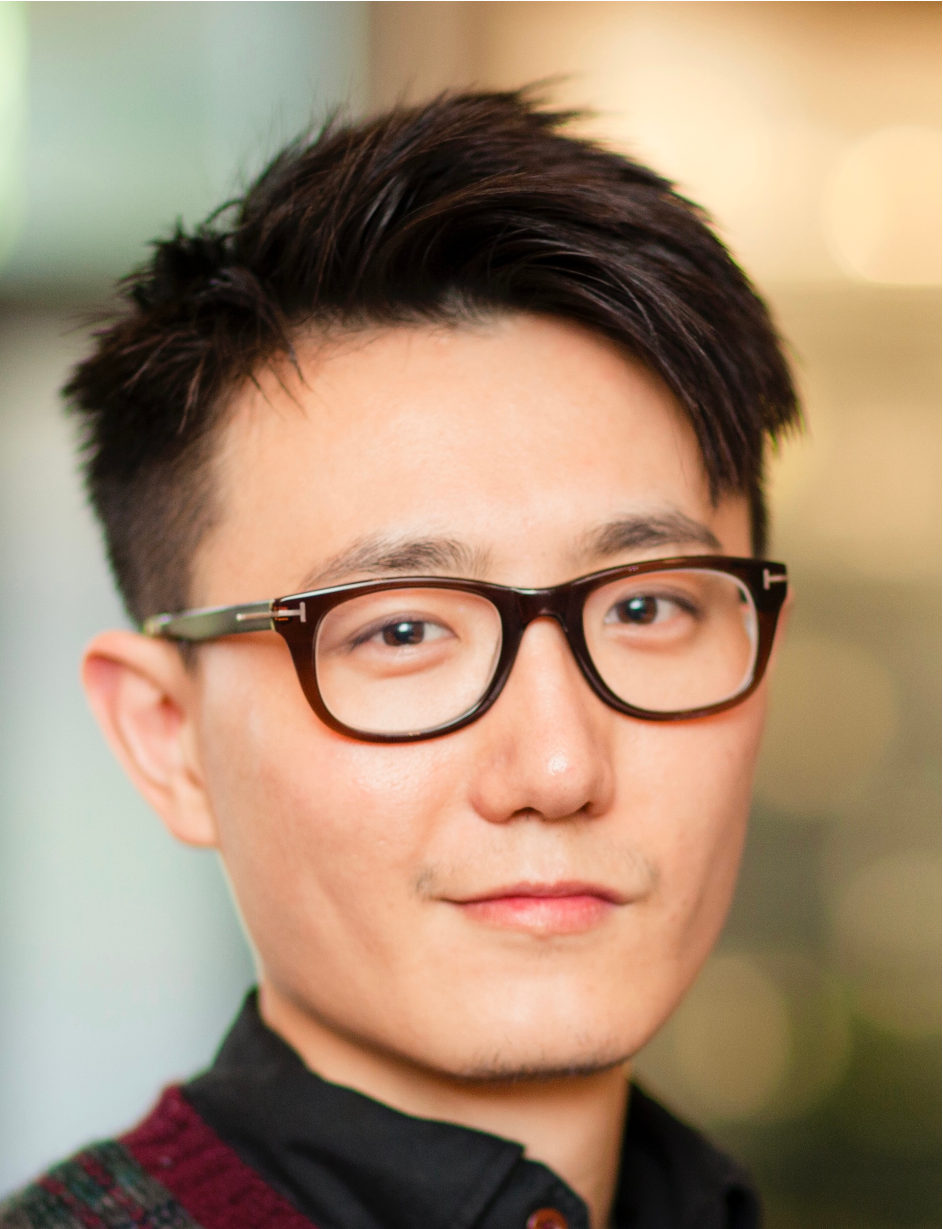}}]{Changlong Wang} is a research fellow at Monash University, specializing in energy system modelling. His Economic Fairways Mapper team was awarded the prestigious Australian Eureka Prize in 2023. Dr Wang is also a Climate Future Fellow at the University of Melbourne and a visiting scholar at the University of Oxford. He represents Australia on multiple IEA Hydrogen TCP tasks on hydrogen modelling.
\end{IEEEbiography}

\vskip -3\baselineskip plus -1fil

\begin{IEEEbiography}[{\includegraphics[width=1in,height=1in,clip,keepaspectratio]{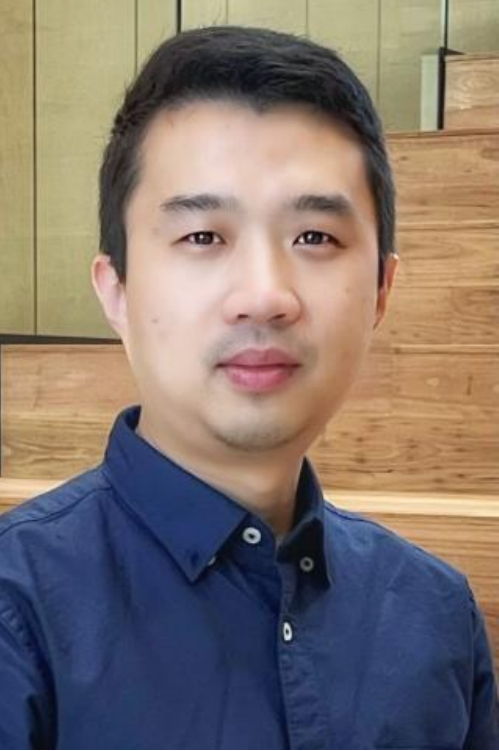}}]{Hao Wang} (M'16) received his Ph.D. in Information Engineering from The Chinese University of Hong Kong in Dec. 2016. He was a Postdoctoral Research Fellow at Stanford University and a Washington Research Foundation Innovation Fellow at the University of Washington. He has been a Senior Lecturer at the Department of Data Science and AI, Faculty of IT, Monash University and is now an ARC DECRA Fellow. His research interests are in optimization, machine learning, and data analytics for power and energy systems.
\end{IEEEbiography}



\end{document}